\documentclass[aps,nofootinbib]{revtex4}

\usepackage{dcolumn}% Align table columns on decimal point
\usepackage{bm}% bold math
\usepackage{amsmath,epsfig,graphicx}

\def\rmd{{\rm d}}
\def\lsim{\mathrel{\rlap{\lower4pt\hbox{\hskip1pt$\sim$}}
\raise1pt\hbox{$<$}}}
\def\gsim{\mathrel{\rlap{\lower4pt\hbox{\hskip1pt$\sim$}}
\raise1pt\hbox{$>$}}}

\newcommand{\erf}[1]{(\ref{#1})} \newcommand{\fl}{\hspace*{-6pc}}

\begin{document}
\title{Semirelativistic approximation to gravitational radiation from encounters with nonspinning black holes}

\author{Jonathan R.\ Gair\footnote{now at: Institute of Astronomy, Madingley Road, Cambridge, CB3 0HA, UK}}
\email{jgair@ast.cam.ac.uk}
\affiliation{Theoretical Astrophysics, California Institute of Technology, Pasadena, CA 91125}

\author{Daniel J.\ Kennefick}
\email{danielk@uark.edu}
\affiliation{Theoretical Astrophysics, California Institute of Technology, Pasadena, CA 91125 and Department of Physics, University of Arkansas, Fayetteville, AR 72701}

\author{Shane L. Larson}
\email{shane@gravity.psu.edu}
\affiliation{Center for Gravitational Wave
Physics, Pennsylvania State University, University Park, PA 16802}

%\recieved{} \accepted{}

\date{\today}

\begin{abstract}
The capture of compact bodies by black holes in galactic nuclei is an
important prospective source for low frequency gravitational wave
detectors, such as the planned Laser Interferometer Space Antenna. This paper calculates, using a semirelativistic
approximation, the total energy and angular momentum lost to
gravitational radiation by compact bodies on very high eccentricity
orbits passing close to a supermassive, nonspinning black hole; these quantities
determine the characteristics of the orbital evolution necessary to
estimate the capture rate.  The semirelativistic approximation
improves upon treatments which use orbits at Newtonian-order and
quadrupolar radiation emission, and matches well onto accurate
Teukolsky simulations for low eccentricity orbits.  Formulae are
presented for the semirelativistic energy and angular momentum fluxes as
a function of general orbital parameters.
\end{abstract}

\maketitle

\section{Introduction}\label{sec:intro}
Proposed space-based gravitational wave interferometers such as the
Laser Interferometer Space Antenna (LISA) will have good sensitivities
in the low frequency gravitational wave band, from about $10^{-4}$ Hz
up to about $1$ Hz.  A promising source of gravitational waves in this
band is the extreme-mass-ratio inspiral (EMRI) of compact objects
(stellar mass black holes, neutron stars, white dwarfs and even main
sequence stars) into massive black holes.  Current estimates suggest
we might detect as many as a thousand EMRI events over the course of
the LISA mission \cite{emri04}.  Gravitational waves from extreme mass
ratio captures will serve as a direct probe of the
innermost population of compact objects around galactic central black
holes, and also provide information on the growth history of such
black holes out to a significant redshift ($z \sim 1$).  In addition
to probing the stellar population of galactic nuclei,
gravitational waves from EMRIs provide a map which encodes the
geometry and structure of the black hole spacetime \cite{ryan97},
allowing a direct comparison of the astrophysical black hole to the
black hole solutions of general relativity.  This technique has been
called ``holiodesy'', in analogy with satellite geodesy, which
observes the motion of small satellites around the Earth to map out
the structure of the planet's gravitational field.

The computation of gravitational radiation from stellar orbits has a
long history.  The classic treatment was that of Peters and Mathews
\cite{PM63,peters64}, who computed the gravitational wave emission
from stars on purely Keplerian orbits in flat space.  The problem of
generation and propagation of gravitational waves in a Kerr background
was addressed by the work of Teukolsky and Press in the early 1970s
\cite{Teuk,PrTeukII,PrTeukIII}, who developed a perturbation formalism
in the Kerr background.  Subsequent work using this formalism has
progressed to the point where the emission from a particle on any
orbit in the Schwarzschild spacetime \cite{CKP} or on a circular
inclined \cite{Hughes00} or eccentric equatorial \cite{GK02} orbit in
the Kerr background can be treated.  Computing the inspiral of stars
on eccentric non-equatorial orbits in Kerr required overcoming some
technical obstacles \cite{Hughes00}, but first results are now
available \cite{drasco04,hughes05}.

In this paper we compute a star's orbital trajectory by
solving the geodesic equations of motion around the black hole, rather
than using Keplerian orbits.  Exact geodesics of the Schwarzschild
spacetime are considered, particularly orbits with high eccentricity,
including marginally bound (parabolic) and unbound (hyperbolic)
orbits.  Here ``parabolic'' is a statement about the orbital energy
$E$ which labels the geodesic trajectory, rather than a statement
about the geometric shape of the orbit in Euclidean geometry. As will be seen in later sections, the orbital trajectories around black holes can exhibit ``zoom-whirl'' behaviour, looping around the black hole more than once ({\it i.e.}, the change
in azimuthal angle $\Delta \phi > 2 \pi$) on any given orbital pass. Even in these situations one can still speak of quantities, such as the eccentricity of the
orbit, which correspond in some formal sense to their Keplerian counterparts.  
Taking these geodesic trajectories as the orbits of the source, we approximate the gravitational radiation using the classic quadrupole formula at Newtonian-order.

This method of obliging the orbiting body to follow a geodesic of the
spacetime, while using the quadrupole approximation to calculate the
gravitational wave emission, has been termed the ``semirelativistic
approximation'' by its originators, Ruffini and Sasaki \cite{RufSas}.
Experience with more accurate, although computationally intensive,
perturbative calculations has shown that when particles make close
approaches to the central black hole, in particular when they are
relatively close to the {\it unstable circular orbit} (UCO) of the
potential (see \ref{geodesics} for a definition), the properties of
the relativistic gravitational potential are of critical importance in
determining the gravitational wave flux.  As will be seen, employing a
more exact description of the particle trajectory (spacetime
geodesics) together with an approximation of the wave flux is much
more accurate in these cases than using a consistent Newtonian-order
approximation.

The semirelativistic approach complements the more complex Teukolsky
based computations in several ways.  First, technical difficulties and
the demands of computing power have made Teukolsky calculations
computationally difficult for orbits with eccentricity near unity, a
regime where the work presented here is designed to work well.
Second, because the computationally intensive Teukolsky approach is
not practical for use in conjunction with typical simulations of the
clusters of stars in galactic centers and their capture rates by the
central black hole, it is useful to look for more convenient
approximate methods, which are sufficiently accurate for reliable
results.

An extension of the semirelativistic approach is currently finding
use in the computation of approximate EMRI waveforms for use in
scoping out LISA data analysis \cite{GHK02,GG05,tev05}.  A particle
inspiral trajectory is computed by integrating post-Newtonian
expressions for the energy and angular momentum fluxes.  Integration
of the Kerr geodesic equations along this trajectory yields the
particle position as a function of time, from which a waveform is
computed from an approximate quadrupole moment tensor generated as in
the semirelativistic approach.  These ``numerical kludge'' waveforms
are inconsistent in that the energy and angular momentum content of
the waveforms differs from the change in energy and angular momentum
of the particle orbit which is nominally emitting the radiation.  This
energy inconsistency means that some of the results that have been
obtained using kludged waveforms, such as signal-to-noise ratios
\cite{emri04}, are inaccurate.  The results of the semirelativistic
calculations presented here allow us to estimate the magnitude of this
inconsistency for inspirals into Schwarzschild black holes (see section~\ref{KludgeInconsis}).  In the future, using semirelativistic fluxes (once these are extended to spinning black holes) in the kludge in place of the post-Newtonian fluxes
employed currently will yield consistent inspirals.

The key results of this paper are summarised as follows:

\begin{enumerate}

    \item Numerical results are presented which enable us to explore
    and evaluate several approximate methods of calculating energy and
    angular momentum fluxes from EMRI orbits, and therefore the
    evolution of those orbits (see section \ref{ELfluxes}).
    Significantly better evolutions (compared to the results of exact
    Teukolsky-based calculations) can be built out of the semirelativistic
    formalism developed here, but improved orbital evolutions can also be
    obtained from classic gravitational radiation estimates (Peters
    and Mathews) simply by choosing to work with ``geodesic
    parameters'' instead of ``Keplerian parameters'', with little
    consequence to computational cost.  See section \ref{compar2} and
    Fig~\ref{PMComp}.

    \item Analytic expressions are derived for the energy ($\Delta E$) and angular
    momentum ($\Delta L_{z}$) radiated in gravitational waves for a
    single orbital pass near a black hole, as a function of the
    orbital parameters, which exactly reproduce our numerical results.
    See section \ref{sec:analexp} and Eqs.\erf{PardeltaE} and
    \erf{PardeltaLz}, for the case of parabolic orbits, and
    appendix~\ref{exactapp} for a more general discussion.

    \item Fitting functions are given which reproduce approximately, but to high
    accuracy, the analytic and numerical results.  These
    functions are relatively simple expressions which could be
    conveniently used in place of consistently Newtonian-order
    expressions such as those of Peters and Mathews, which are
    significantly less accurate for orbits with very close periapses
    (see section \ref{fitfunc} and appendix~\ref{fitapp}).  Although
    we present fits to the semirelativistic results only, the fitting
    functions have more general applicability and it should be
    possible to derive a fit of the same form to Teukolsky-based
    results once these are available for generic orbits.

\end{enumerate}

The remainder of the paper will be organised as follows.  In section
\ref{sec:qandp} we describe the semirelativistic scheme which we use
to model the gravitational radiation from EMRIs.  In section \ref{ELfluxes} we present 
fluxes calculated using this approach and compare these with more
accurate Teukolsky-based results, as well as with the consistently Newtonian results of Peters and Mathews.  We also present
analytic formulae which reproduce our numerical results, and discuss
the case of hyperbolic orbits which are initially unbound but become bound to the black hole via gravitational bremsstrahlung.  Finally in section \ref{sec:Discussion} we summarise our most important results and findings.

Throughout this paper, geometric units where $G = c = 1$ are employed
unless otherwise specified.

\section{Model of gravitational radiation}\label{sec:qandp}

\subsection{Quadrupole approximation}
The energy and angular momentum carried away by gravitational waves
from a weak-field, slow-motion source can be computed using the
quadrupole formula \cite{mtw}
\begin{equation}
    \frac{dE}{dt} = - \frac{1}{5} \langle \dddot{{\cal I}}_{jk}
    \dddot{{\cal I}}_{jk} \rangle \ ,
    \label{QuadEluminosity}
\end{equation}
and
\begin{equation}
    \frac{dL_{z}}{dt} = - \frac{2}{5} \epsilon_{jkl} \langle
    \ddot{{\cal I}}_{ka} \dddot{{\cal I}}_{al} \rangle \ ,
    \label{QuadLzluminosity}
\end{equation}
where summations are implied over repeated indices, $\epsilon_{jkl}$
is the permutation symbol, and ${\cal I}_{jk}$ is the reduced
quadrupole moment of the system,
\begin{equation}
    {\cal I}_{jk} = \int \rho (x_{j} x_{k} - \frac{1}{3}
    \delta_{jk}r^{2})d^{3}x\ .
    \label{reducedQM}
\end{equation}
The angle brackets $\langle\,\rangle$ in
\erf{QuadEluminosity}--\erf{QuadLzluminosity} indicate averaging over
several orbits, but parabolic trajectories (our main focus) do not
have periodic orbits.  Indeed, the period of a parabolic orbit is
infinite, so the average energy flux over the whole orbit is zero.
Therefore it is convenient to work instead in terms of $\Delta E$ and
$\Delta L_{z}$, the total energy and angular momentum radiated over a
single orbit, which are in general finite.

The corresponding gravitational waveform in transverse-traceless gauge
is given by \cite{mtw}
\begin{equation}
h_{jk}^{TT}=\frac{2}{r}\, \left[ P_{jl}\,\ddot{{\cal
I}}_{lm}\,P_{mk}-\frac{1}{2}\,P_{jk}\,P_{ml}\,\ddot{{\cal
I}}_{lm}\right], \qquad P_{jk}=\delta_{jk}-n_{j}\,n_{k},
\label{QuadWave}
\end{equation}
in which $n_{j}$ denotes the direction of propagation of the wave and
$r$ is the proper distance to the source.

For orbits in the weak-field, far from the black hole, the quadrupole
formula applies, the source particle orbit is Keplerian, and the
radiation field reduces to the Peters and Mathews result.  For orbits
that pass close to the black hole, the particle's geodesic orbit is no
longer Keplerian and the motion is neither weak-field nor slow motion,
so the quadrupole formula does not describe the wave emission
precisely.  As described above, correcting the emission formula
requires use of black hole perturbation theory (Teukolsky methods),
which is computationally very challenging, but a significantly
improved approximation can be obtained by using the quadrupole formula
with source particle orbits modified to be a geodesic of the black
hole spacetime.

To do this, first identify the Cartesian coordinates, $x_{j}$, in the
quadrupole moment expression \erf{reducedQM} with coordinates in the
Schwarzschild spacetime.  Treating the Schwarzschild coordinates $\{r,
\theta, \phi\}$ as spherical polar coordinates, define a set of
pseudo-Cartesian coordinates by
\begin{equation}
    x^{i} = \left( r \sin \theta \cos \phi, r \sin \theta \sin \phi, r
    \cos \theta \right)\ .
    \label{particleCoords}
\end{equation}
In these coordinates, one can solve the Schwarzschild geodesic
equations (see section \ref{geodesics}) to compute the particle orbit
$x_{j}(t)$, and substitute the resulting trajectory into the
quadrupole moment expression \erf{reducedQM} to compute
\begin{equation}
    {\cal I}^{jk} = m \left[ x^{j}x^{k} - \frac{1}{3} \delta^{jk}
    r^{2} \right]\ ,
    \label{Qjk}
\end{equation}
where $m$ is the mass of the particle.  Finally, we compute an
estimate of the energy and angular momentum radiated using expressions
\erf{QuadEluminosity}--\erf{QuadLzluminosity}.

This approximation for gravitational wave emission was first applied
by Ruffini and Sasaki, who termed it a ``semirelativistic
approximation'' \cite{RufSas}, since it makes use of the fully
relativistic orbit, but only a weak-field expression for the
gravitational waves.  The approach is equivalent to attaching the
compact body to a string in flat space and forcing it to move on a
path that corresponds to a geodesic of the Schwarzschild potential.
In reality, the inspiralling body does not follow a geodesic, due to
the effect of radiation reaction on the orbit. The loss of energy
occurs continuously, so particle trajectories depart from a true
geodesic path continuously.  Instead of stable orbits, particles
follow inspiralling paths with a steadily decreasing average radial
distance from the center.  However, in the typical case for extreme
mass ratio inspirals, in which the rate of energy loss per orbit is
small, the actual particle trajectory looks similar to a geodesic
orbit for long periods, so one is justified in making an
adiabatic approximation \cite{Hughes00}; simply assume the body
evolves through a sequence of geodesics and determine this sequence
using the energy and angular momentum fluxes from each geodesic orbit.

The adiabatic approximation will break down when the orbital 
parameters change significantly on the timescale of a single orbit,
which occurs only very close to the final plunge. The trajectory and
waveform in this region must be computed using the computationally
intensive self-force formalism (see \cite{poisson04} for a review).

\subsection{Geodesics}\label{geodesics}
The equations governing geodesic motion in the Schwarzschild
spacetime, in the usual Schwarzschild coordinates, are given by
\begin{eqnarray}
   \left(\frac{\rmd r}{\rmd\tau}\right)^2 &=& \left( E^{2} - 1\right)
   +\frac{2\,M}{r}\,\left(1+\frac{L_{z}^{2}}{r^{2}}\right)
   -\frac{L_{z}^{2}}{r^{2}} \,, \label{rdot} \\
   r^{2}\,\left(\frac{\rmd\phi}{\rmd\tau} \right) &=& L_{z} \,,
   \label{phidot} \\
   \left(1-\frac{2\,M}{r}\right)\,\left(\frac{\rmd t}{\rmd\tau}
   \right) &=& E ,  \label{tdot}
\end{eqnarray}
where $\tau$ is the proper time along the geodesic, $L_{z}$ is the
conserved specific angular momentum of the particle, $E$ is the
conserved specific energy and $M$ is the mass of the central black
hole.  We have taken advantage of spherical symmetry to assume an
equatorial orbit, $\theta=\pi/2$, without loss of generality.  The
radial equation of motion \erf{rdot} may be written as a cubic
polynomial divided by $r^3$.  The cubic has one, two or three roots
depending on the values of $E$ and $L_{z}$.  These roots correspond to
turning points of the radial motion.  Orbits with a single turning
point plunge into the black hole and correspond to energies
\begin{eqnarray}
E^{2}-1 &>& \frac{L_{z}^{2}}{r_{I}^{2}} - \frac{2\,M}{r_{I}} -
\frac{2\,M\,L_{z}^{2}}{r_{I}^{3}}, \label{Emin} \\
{\rm where} & & r_{I} =
\frac{L_{z}^{2}}{2\,M}\,\left(1-\sqrt{1-\frac{12\,M^{2}}{L_{z}^{2}}}
\right) \nonumber
\end{eqnarray}
For bound orbits, it is possible to define an orbital eccentricity by
analogy with the Keplerian case.  Define the position of the apoapse
of the orbit to be
\begin{equation}
   r_{a} = \frac{1+e}{1-e} \, r_{p} ,
   \label{eccdef}
\end{equation}
where $r_{p}$ is the radius of the periapse.  Equation \erf{eccdef} is
used to define the eccentricity of a geodesic in terms of the turning
points of the orbit \cite{CKP,GK02}.  This definition carries over to
parabolic ($E^2=1$) and hyperbolic orbits ($E^{2}>1$).  In the
parabolic case, the radial geodesic equation \erf{rdot} has only two
turning points (the apoapse is `at infinity'), but definition
\erf{eccdef} holds with $e=1$.  In the hyperbolic case, one of the
turning points has $r<0$; using this in \erf{eccdef} one finds $e > 
1$, and so in this case we call that turning point the apoapse.

For this definition for eccentricity, we use the parameters
$(r_{p},e)$ to characterise the orbit, instead of $(E,L_{z})$.  The
energy and angular momentum are related to the periapse and
eccentricity by
\begin{eqnarray}
    E &=& \sqrt{1+\frac{M\,(1-e)(4\,M-(1+e)r_{p})}{r_{p}((1+e)\,
    r_{p}-(3+e^{2})\,M)}}\,, \label{Eofrpe} \\
   L_{z} &=& \frac{(1+e)\,r_{p}}{\sqrt{(1+e)\,
   \frac{r_{p}}{M}-(3+e^{2})}} \,.
   \label{Lofrpe}
\end{eqnarray}
The radial geodesic equation becomes
\begin{equation}
   \left(\frac{\rmd r}{\rmd\tau}\right)^2 = \left(E^{2}-1\right) \,
   \left(\frac{1}{r}\right)^{3}\,\left(r_{a} - r
   \right)\left(r-r_{p}\right)\left(r-r_{-}\right) \label{rdotrpe}
\end{equation}
where the apoapse, $r_{a}$, and energy $E$ are given by equations
\erf{eccdef} and \erf{Eofrpe}, and the third root of the potential is
given by
\begin{equation}
   r_{-} = \frac{2(1+e)\,r_{p}}{((1+e)r_{p}-4\,M)} \label{rminus} \,.
\end{equation}
For any given eccentricity, there is a minimum value for the periapse
below which the orbit plunges directly into the black hole.  This
occurs when the two inner turning points of the geodesic equation,
$r_{-}$ and $r_{p}$, coincide.  A geodesic with precisely this
periapse asymptotically approaches a circular orbit as it nears the
periapse, and spends an infinite amount of time whirling around the
black hole.  The asymptotic circular orbit is an unstable orbit of the
gravitational potential, and we will refer to it as the `unstable
circular orbit' (UCO).  The radius of the UCO determines the minimum
periapse for geodesics of a fixed eccentricity.  Equating $r_{-}$ and
$r_{p}$ yields an expression for the UCO in terms of $e$
\begin{equation}
   r_{UCO} = \frac{2\,(3+e)}{1+e} \, M. \label{rUCO}
\end{equation}
The statement that orbits with $r_{p} < r_{UCO}$ are plunging is
equivalent to the relationship \erf{Emin} between the energy and
angular momentum (see \cite{CKP} for an equivalent relation in terms
of the semi-latus rectum).  If $e=0$ the UCO is at the familiar innermost stable circular orbit, $r=6M$.  In the extreme hyperbolic limit, $e \rightarrow \infty$, the
UCO approaches the horizon $r=2M$.  Parabolic orbits ($e=1$) have a
minimal periapse of $r=4M$.  Cutler, Kennefick and Poisson \cite{CKP}
also discuss the UCO, but they call the line $r_{p}=r_{UCO}$ the
`separatrix', since it separates bound from plunging orbits in phase
space.

\begin{figure}
\centerline{\includegraphics[keepaspectratio=true,height=4in,
			     angle=0]{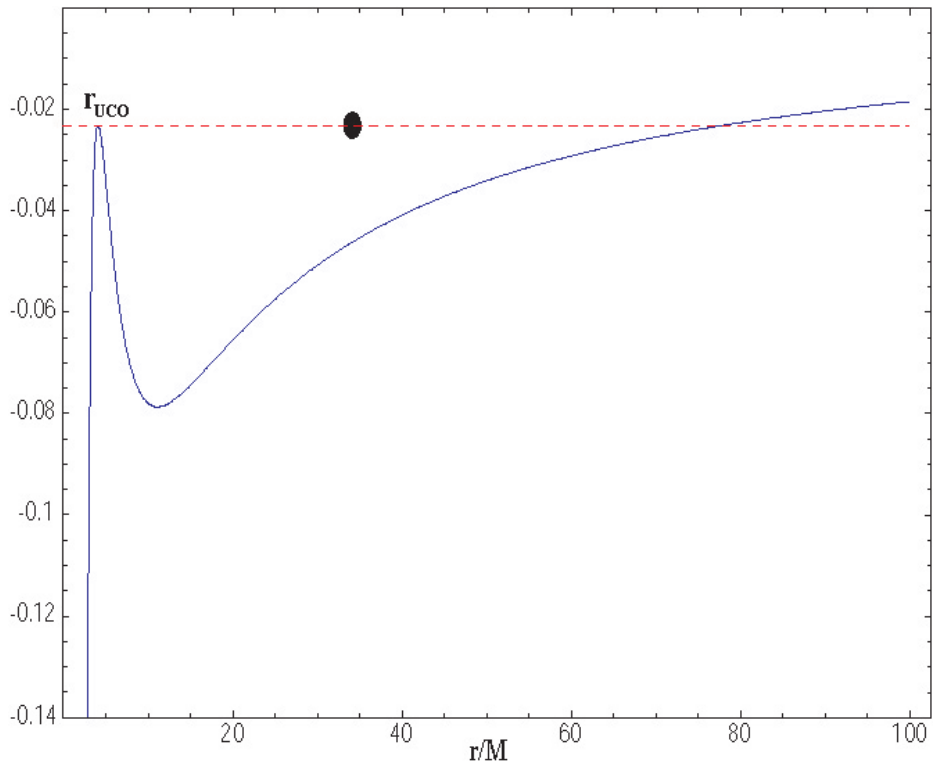}}
\caption{Radial gravitational potential for
a zoom-whirl orbit.  The dashed line corresponds to the energy of
the orbit.  The orbit oscillates in the region where the potential
(solid curve) lies below the energy line.  If the energy is too high
and the orbit passes inside the maximum of the radial potential
($r_{UCO}$), the particle plunges into the black hole.}
\label{ZWpot}
\end{figure}

These orbital properties can be understood by considering the radial
gravitational potential $V(r,L_{z})$, which is illustrated in
Figure~\ref{ZWpot} for a typical zoom-whirl orbit.
%shown in Figure~\ref{ZWorbit}. 
The characteristic feature of these highly
relativistic potentials is the maximum at $r_{UCO}$.  As an inspiral
approaches plunge, the orbit is within the potential well but close to
the top of the well.  That is, the periapse lies close to the UCO.
The particle thus zooms out to apoapse and back, but loiters close to
periapse, whirling several times around the black hole on a nearly
circular orbit before zooming out to apoapse again.  As one approaches
the UCO, the more exaggerated the whirl phase gets and the closer the
resemblance to an unstable circular orbit.

\subsection{Waveform structure}
\label{sec:Waveforms}
\begin{figure}
\centerline{\includegraphics[keepaspectratio=true,height=3.75in,
			     angle=0]{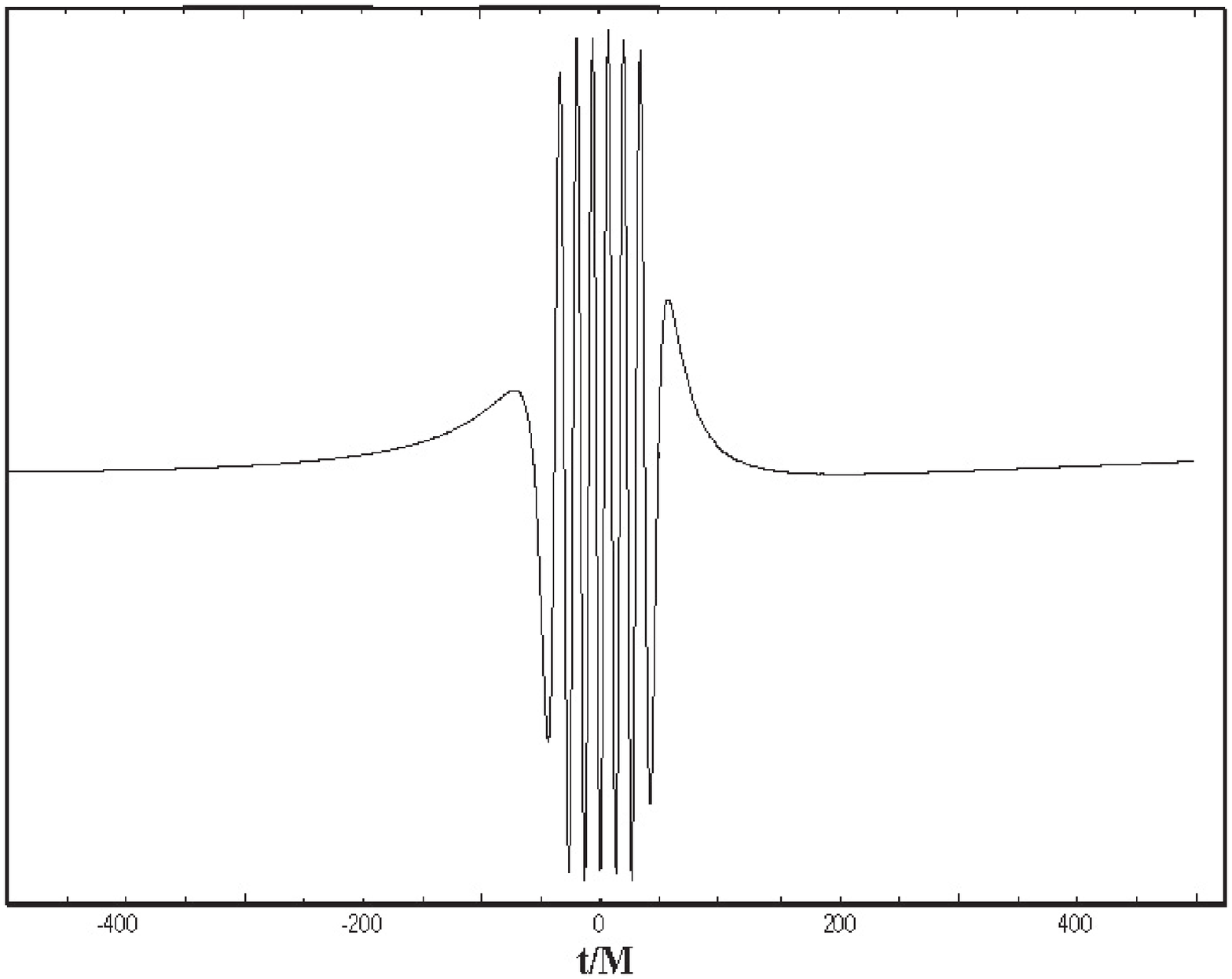}}
\vspace{0.25in}
\centerline{\includegraphics[keepaspectratio=true,height=4.75in,
			     angle=-90]{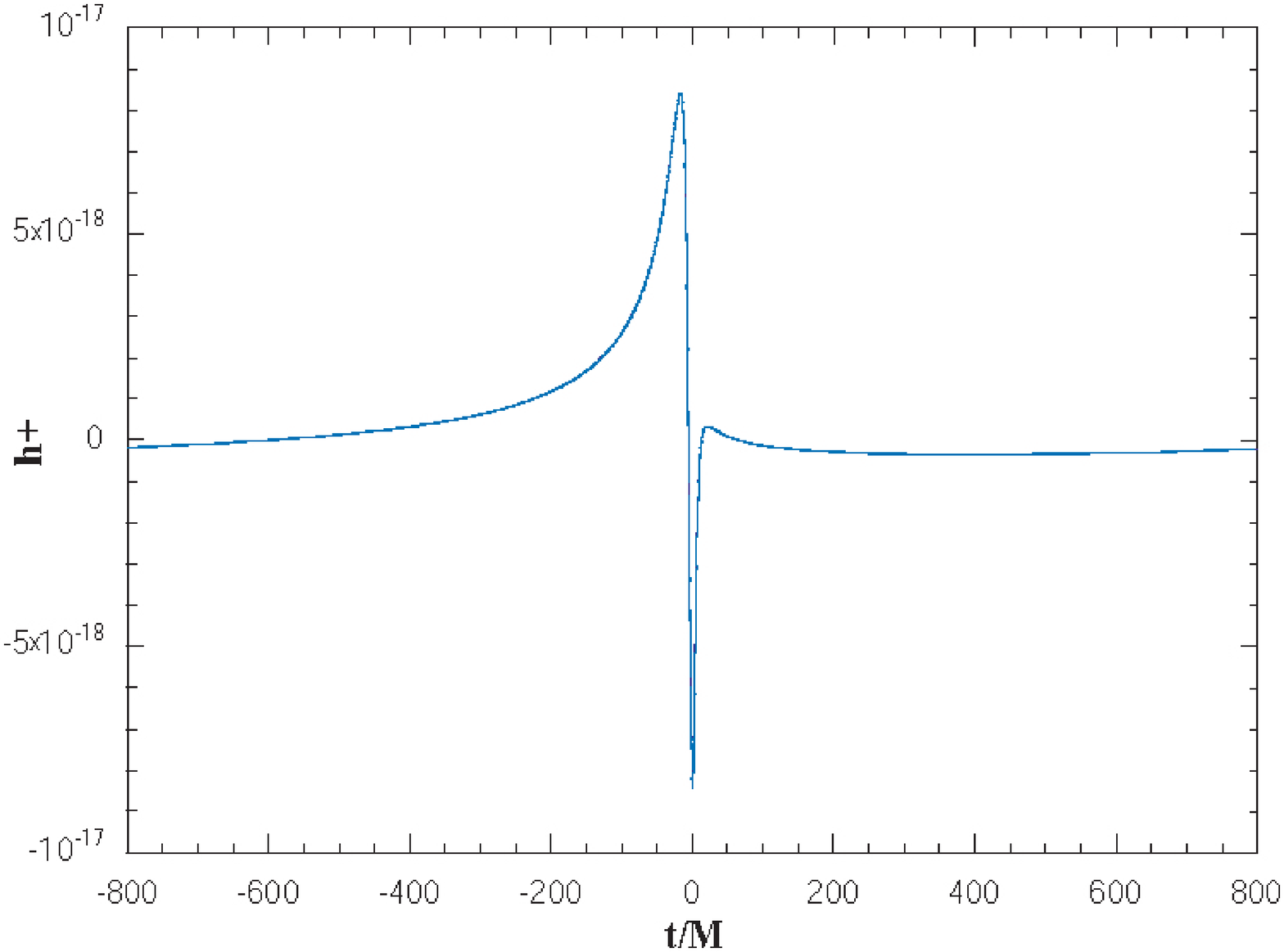}}			     
\caption{Sample gravitational waveform (top) from the zoom-whirl orbit
indicated in Figure~\ref{ZWpot}.  We show the plus polarisation of
the gravitational wave as a function of time.  The radiation is
emitted predominantly in a high frequency burst during the whirl phase
of the orbit. For comparison, the waveform from the Keplerian orbit 
with the same parameters is shown in the bottom diagram.}
\label{ZWwave}
\end{figure}

The waveforms resulting from zoom-whirl orbits are easy to
comprehend.  During the long apoapse passage the motion of the source
is relatively slow, and the amplitude and frequency of the
gravitational wave produced are both low.  Near periapse the motion
is much more rapid and the signal has much higher amplitude and
frequency.  If the whirling phase of the orbit consists of one or more
complete revolutions about the central body then the waveform will
have several cycles (two for each revolution).  The result is a
waveform with a very low frequency (determined by the radial period of
the orbital motion) and low amplitude superposed with a burst of short
duration (relative to the overall period) and relatively high
amplitude whose frequency is determined by the azimuthal period of the
orbital motion.  An example waveform is shown in Figure~\ref{ZWwave},
corresponding to the orbit indicated in Figure~\ref{ZWpot}.

Note that while the radial frequency is much too low for detection by
LISA, the azimuthal ($\phi$) frequency does fall in the LISA bandwidth
for the orbits of interest.  Although there is a low probability of
detecting these bursts since they are too brief and infrequent
(typically occurring once every few years or even longer, depending on
the radial period) to have high signal-to-noise, the background of all
such bursts occurring throughout our neighbourhood of the Universe will
create an astrophysical background of noise from which other sources
must be subtracted \cite{barack04}.

\section{Energy and Angular Momentum Fluxes}\label{ELfluxes}
The semirelativistic approximation is constructed by integrating
approximate rates of energy and angular momentum flux over
relativistically accurate geodesic orbits.  A consistent approximation
would require that the particle orbit be approximated to the same
level of accuracy as employed for the fluxes.  A Newtonian-order
approximation, such as that of Peters and Mathews, makes use of
Keplerian elliptical orbits in flat spacetime and the fluxes are of
quadrupole order only.  There might not appear to be much
justification for using accurate orbital paths but retaining
approximate fluxes.  For orbits which never come close to the central
black hole the semirelativistic scheme does not improve significantly
on fully consistent Newtonian approximations \--- in nearly flat
regions of spacetime all reasonable approximations fare well.
However, orbits with small periapse distances are a very different
case.  More accurate schemes (such as those based on solution of the
Teukolsky equation) show that the radiation from orbits which come
close to the black hole show features that are greatly modified by the
strongly curved spacetime and which are qualitatively different from
those seen at large radii from the black hole.  Many of these features
arise from the properties of the geodesics in the strong-field regime
and therefore, as argued in \cite{GHK02}, such features can be
modelled by schemes which combine exact geodesics with approximate
fluxes.  This approach shows significant improvements over the weak
field approximation \cite{PM63}, as we will see in the next section,
while being considerably less expensive computationally than solving
the Teukolsky equation.

\subsection{Comparison to Peters and Mathews and Teukolsky results}
\label{compar2}
The expressions derived in \cite{PM63} for the energy and angular
momentum fluxes from a Keplerian orbit are
\begin{eqnarray}
    \left<\frac{\rmd E}{\rmd t}\right> &=& -\frac{32}{5} \,
    \frac{m}{M^{2}} \, \frac{(1-e)^{\frac{3}{2}}}{\left(1+
    e\right)^{\frac{7}{2}}}\,\left(1+\frac{73}{24}\,e^{2}+
    \frac{37}{96}\,e^{4}\right)\, \left(\frac{r_{p}}{M}\right)^{-5}
    \label{PMdEdt} \\
    \left<\frac{\rmd L_{z}}{\rmd t} \right> &=& 
-\frac{32}{5}\,\frac{m}{M}\,
    \frac{(1-e)^{\frac{3}{2}}}{\left(1+e\right)^{2}}\,
    \left(1+\frac{7}{8}\,e^{2}\right)\,\,
    \left(\frac{r_{p}}{M}\right)^{-\frac{7}{2}} \,.  \label{PMdLdt}
\end{eqnarray}
These are equations (5.4) and (5.5) of \cite{peters64}, but written 
to lowest order in the mass ratio, $m/M$, in the extreme-mass-ratio 
limit 
$M=m_{1}\gg m_{2}=m$. The eccentricity and periapse of a Keplerian 
orbit are given in terms
of the energy and angular momentum by
\begin{equation}
   e^{K} = \sqrt{1-\frac{L_{z}^{2}}{M^{2}}\,\left(1-E^{2}\right)}, 
\qquad 
   r_{p}^{K} = \frac{L_{z}^{2}}{M\,(1+e^{K})} \label{erpKep}
\end{equation}
To use \erf{PMdEdt}--\erf{PMdLdt} in the strong-field regime, the
natural way to proceed is to evaluate the fluxes in equations
\erf{PMdEdt}--\erf{PMdLdt} for the Keplerian orbit with the
corresponding energy and angular momentum, i.e., substitute $e^{K}$
and $r_{p}^{K}$ from \erf{erpKep} into \erf{PMdEdt}--\erf{PMdLdt}
(``Peters and Mathews with Keplerian parameters'').  This approach
runs into difficulties however, since Keplerian orbits do not exist
for certain valid choices of $E$ and $L_{z}$, for example if
$L_{z}^{2} > M^{2}/(1-E^{2})$ the Keplerian eccentricity is undefined.
An alternative way to proceed is to compute the geodesic eccentricity
and periapse using expressions \erf{Eofrpe}--\erf{Lofrpe} and use
these in the flux formulae \erf{PMdEdt}--\erf{PMdLdt}, thus
identifying geometrically similar orbits (``Peters and Mathews with
geodesic parameters'').

\begin{figure}
%\centerline{\psfig{file=tPSGbytGW_e05.eps,width=8cm,angle=0}}
\includegraphics[keepaspectratio=true,height=3.55in,
		 angle=0]{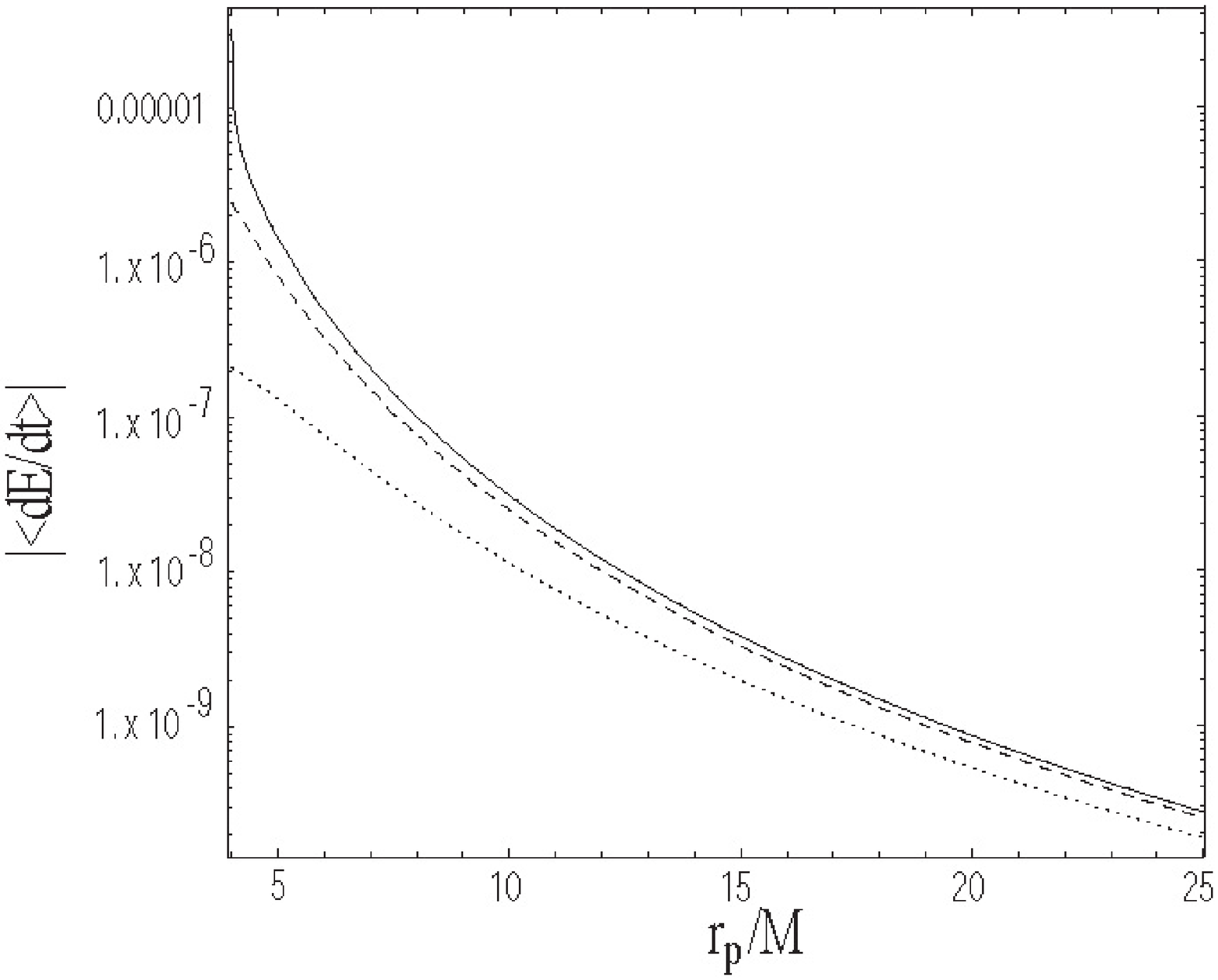}

%\centerline{\psfig{file=tPSGbytGW_e099.eps,width=8cm,angle=0}}
\includegraphics[keepaspectratio=true,height=3.55in,
		 angle=0]{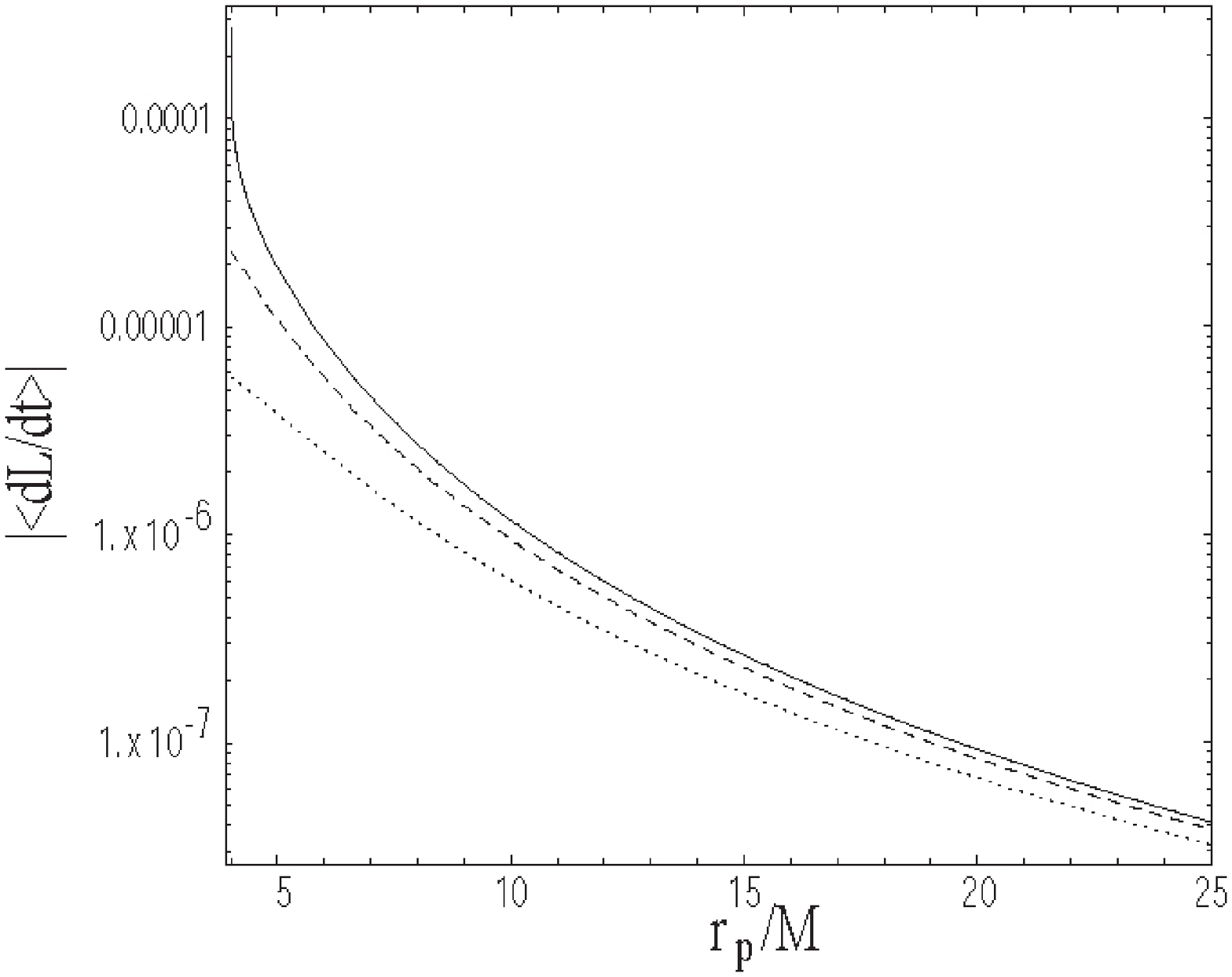}

\caption{Comparison between the semirelativistic results and Peters
and Mathews as a function of periapse for orbits with fixed geodesic
eccentricity $e=0.99$.  The solid lines are the results from the
semirelativistic approximation discussed here.  The dashed and dotted
lines are the Peters and Mathews results with geodesic parameters and
Keplerian parameters respectively.  We use a logarithmic vertical
scale and plot the absolute value of the energy (upper plot) and
angular momentum (lower plot) fluxes.}
\label{PMComp}
\end{figure}

In Figure~\ref{PMComp} we compare the fluxes computed in these three
ways: Peters and Mathews fluxes using Keplerian parameters, Peters and
Mathews using geodesic parameters, and the semirelativistic
approximation, all as a function of geodesic (relativistic) periapse
for fixed geodesic (relativistic) eccentricity of $e=0.99$.  For large
periapse, the three approximations agree as expected, but once the
periapse becomes moderate ($r_{p} \lsim 50M$), the Peters and Mathews
expression with Keplerian parameters begins to differ quite
significantly from the other approximations.  In the strong-field
region ($r_{p} \lsim 10M$), the semirelativistic approximation begins
to differ significantly from both applications of the Peters and
Mathews formula, predicting greater fluxes of both energy and angular
momentum.

\begin{figure}
\centerline{\includegraphics[keepaspectratio=true,width=3.35in,
			     angle=-90]{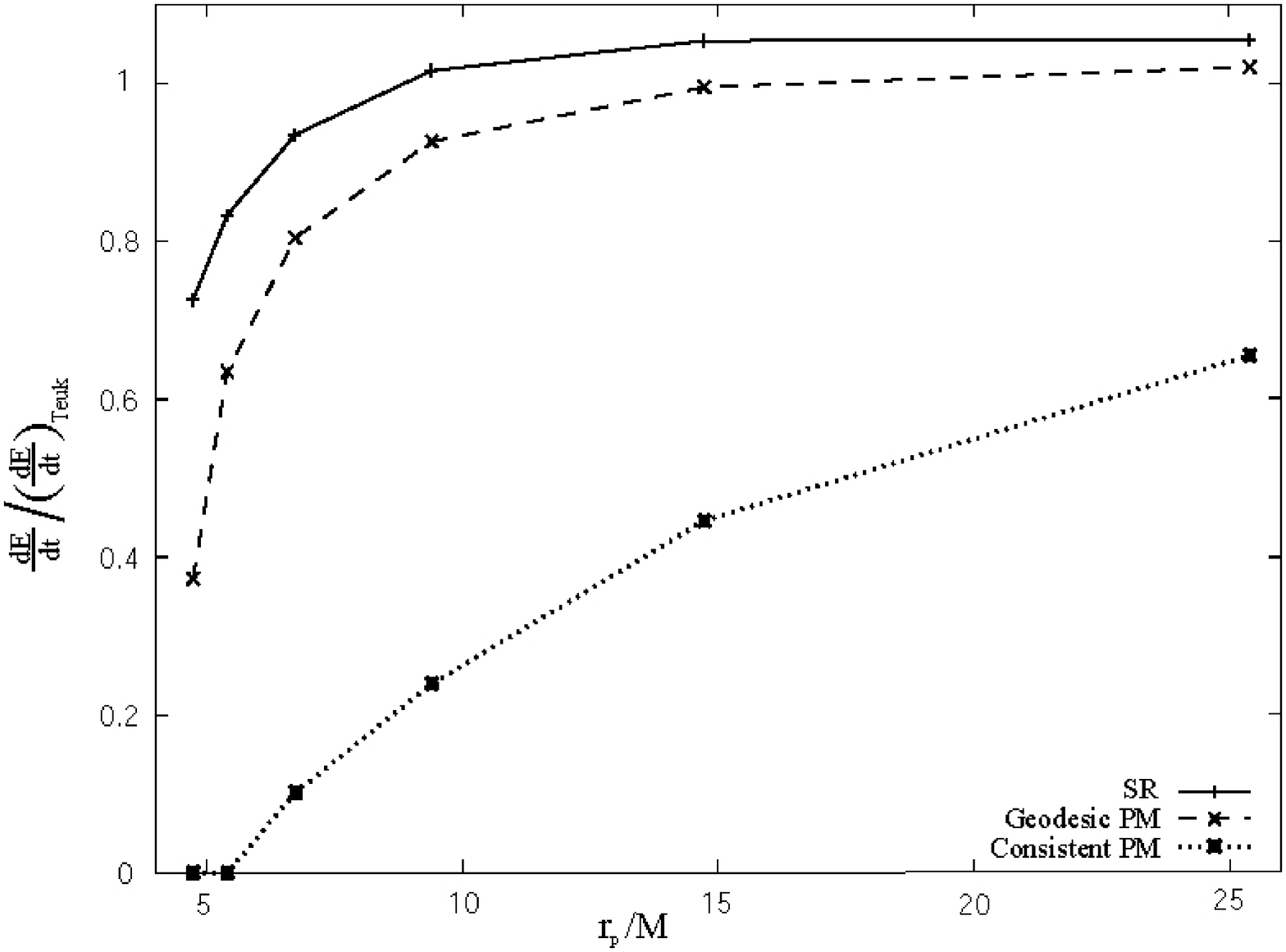}}
\centerline{\includegraphics[keepaspectratio=true,width=3.35in,
			     angle=-90]{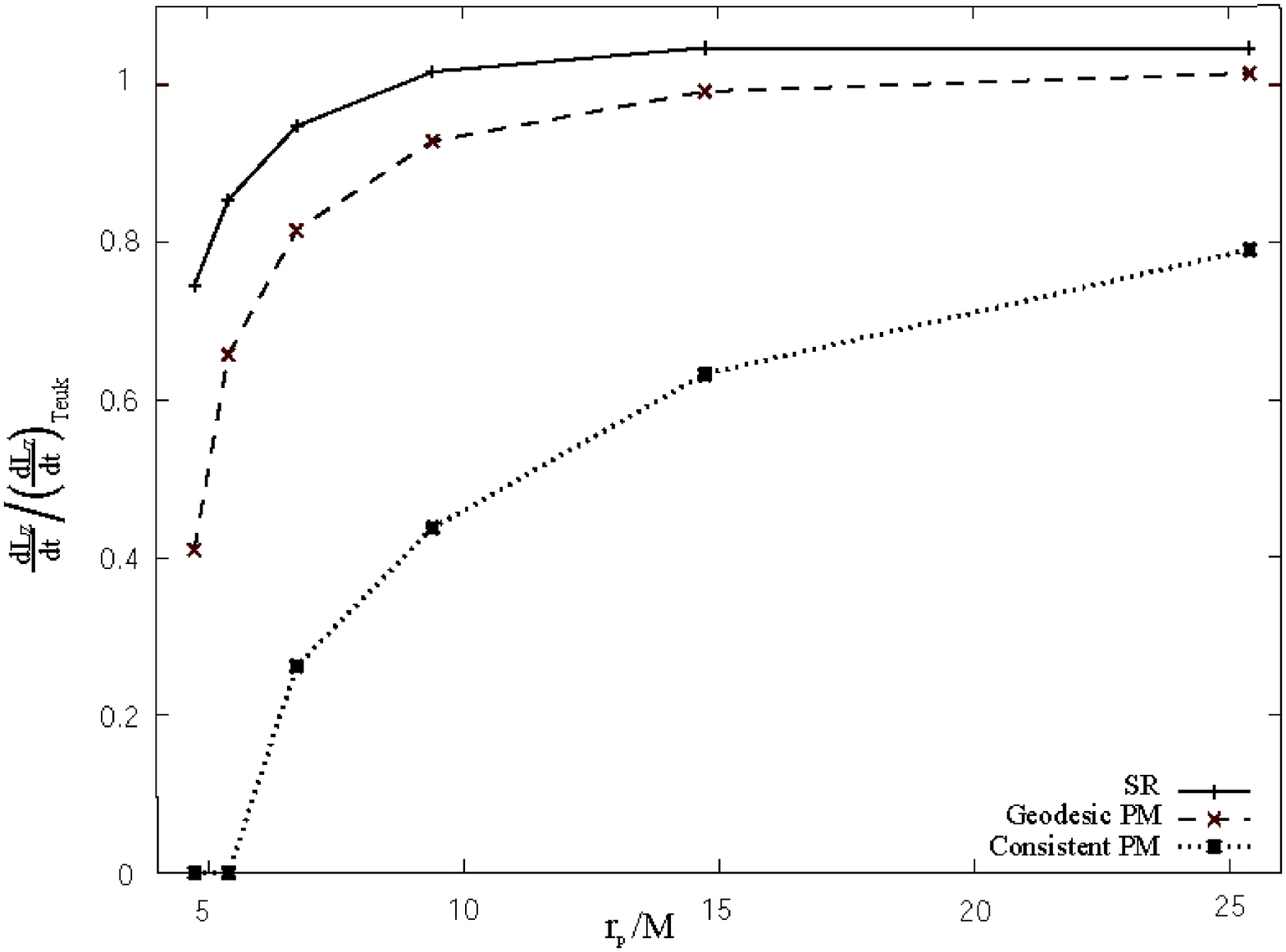}}
\caption{Comparison between accurate Teukolsky results and various 
approximations, for orbits with fixed eccentricity
$e=0.5$ and a variety of periapses. The plots show the ratio of the 
flux computed using a given approximation to the flux obtained from 
the Teukolsky calculation. The upper plot shows the ratio of the 
energy fluxes, while the lower plot is the ratio of the angular 
momentum fluxes. In both plots, the solid lines are the 
semirelativistic results.  The dashed and dotted lines are for
Peters and Mathews, evaluated with geodesic parameters and Keplerian 
parameters, respectively.  The latter lines cut off at small periapse 
since there are no corresponding Keplerian orbits in that region.}
\label{TeukComp}
\end{figure}

To verify that the semirelativistic results are an {\it improvement}
over Peters and Mathews, rather than merely being {\it different}, the
approximation can be compared to perturbative results from integration
of the Teukolsky equation.  Very few results are available for high
eccentricities in the Teukolsky formalism, so the comparisons here are
shown at lower eccentricities.  In Figure~\ref{TeukComp}, the
semirelativistic and Peters and Mathews fluxes are compared to
Teukolsky calculations \cite{GK02} for orbits with eccentricity $e=0.5$ and a
variety of periapses.  As one would expect, the semirelativistic
approximation is not superior to a consistent Newtonian approach for
periapses greater than about $\sim 50M$ (sometimes it does worse and
sometimes better than the Peters and Mathews results, but never
extremely different).  For periapses less than $\sim 50M$, the
semirelativistic method improves significantly upon the consistent
Peters and Mathews approximation.  The improvement gained using the
semirelativistic approximation was also noted by \cite{Tanaka} in
comparisons to a selection of Teukolsky-based results.  Thus for the
type of orbit of interest to this paper (highly eccentric orbits with
close periapses), consistent Newtonian approximations should be
regarded with suspicion, but approximations which make use of exact
geodesics (like the semirelativistic approximation), will fare very
well qualitatively and quite well quantitatively, as long as the
periapse is not extremely small.

What is perhaps more surprising is that one may obtain an improved
approximation from the Newtonian-order expressions (i.e., Peters and
Mathews) if one carefully chooses the Newtonian parameters which are
to be equated with the ``true'' curved space parameters (i.e.,
geodesic parameters rather than Keplerian parameters).  While the
semirelativistic approximation is always an improvement over this in
the strong-field regime, the gain is only significant for very close
periapses, $r_{p} \lsim 10 M$.  In fact, for small eccentricities
there is no significant gain using the semirelativistic fluxes.  For
a circular orbit of radius $r_{0}$, the quadrupole formulae
\erf{QuadEluminosity}--\erf{QuadLzluminosity} tell us that $\langle
\rmd E/\rmd t \rangle = -32\,r_{0}^{4}\,\Omega_{\phi}^{6}$ and
$\langle \rmd L_{z}/\rmd t \rangle =
-32\,r_{0}^{4}\,\Omega_{\phi}^{5}$, where $\Omega_{\phi} = \rmd
\phi/\rmd t$ is the angular velocity.  For both a Keplerian orbit and
a circular geodesic of the Schwarzschild metric,
$\Omega_{\phi}=1/r_{0}^{\frac{3}{2}}$.  Therefore, the standard Peters
and Mathews result is recovered exactly for circular orbits using
either geodesic or Keplerian parameters.

We are primarily interested in highly eccentric orbits, for which the
semirelativistic results are a significant improvement over {\it any}
method based on Peters and Mathews.  Nonetheless, if one does not wish
the additional computational burden of evaluating more accurate
semirelativistic flux expressions, a significant improvement can
still be gained by evaluating the Newtonian fluxes using geodesic
parameters.

\subsection{Phase space structure}
An inspiral sequence may be constructed from the semirelativistic
fluxes by integrating ($\rmd E/\rmd t$, $\rmd L_{z}/\rmd t$).  While
the duration of the inspiral depends on the value of $\rmd E/\rmd t$,
the sequence of geodesics that the inspiral passes through in the $(E,
L_{z})$ phase space depends only on the ratio $\rmd E/\rmd L_{z}$.
Equivalently, in the $(r_{p}, e)$ phase space, it depends only on
\begin{equation}
   \label{drpde}
   \frac{\rmd r_{p}}{\rmd e} = \frac{\frac{\partial L}{\partial
   e}\,\frac{\rmd E}{\rmd L_{z}}-\frac{\partial E}{\partial
   e}}{\frac{\partial E}{\partial r_{p}}-\frac{\partial L_{z}}
   {\partial r_{p}}\,\frac{\rmd E}{\rmd L_{z}}}
\end{equation}
It turns out that the semirelativistic approximation reproduces the
ratio $\rmd E/\rmd L_{z}$ to a very high accuracy when compared to the
Teukolsky results.  While the value of $\Delta E$ can differ by as
much as $25 \% $, the ratio $\rmd E/\rmd L_{z}$ is recovered to better
than $5\% $.  This means that the structure of the semirelativistic
phase space will be a good approximation to the true structure,
although there is some error in the estimated duration of inspirals.

This is an interesting result from the point of view of detection of
EMRIs with LISA. An error in the timescale of an inspiral can be
partially corrected by a change in the mass ratio.  Since the phase
space trajectory is well approximated, an inspiral waveform computed
using this approach might be a moderately good fit to a true inspiral
waveform with a slightly different mass ratio, and therefore could be
used as a detection template over sufficiently short time stretches.
This may not be practical, since the error in $\rmd E/\rmd t$ is not a
constant factor, which would require varying the mass ratio over the
inspiral.  Moreover, other features that these approximations do not
include (such as the ``conservative'' part of the self-force) may lead
to rapid de-phasing of the kludge templates.  This is nonetheless an
interesting result.

The accuracy with which the phase space structure is reproduced can be
understood by considering what happens at the extremes of an inspiral.
When the periapse is very large, the semirelativistic approximation
is good, and is expected to reproduce $\rmd E/\rmd L_{z}$ accurately.
For an orbit that sits on the separatrix between plunging and
non-plunging orbits ($r_{p} \sim r_{UCO}$), the geodesic spends an
infinite amount of time whirling about the black hole on a nearly
circular orbit at the UCO. The flux of energy and angular momentum is
totally dominated by the circular part of the orbit.  For any
radiation field in a spherically symmetric spacetime, the energy and
angular momentum fluxes carried away from a circular orbit obey the
relation $dE/dt = \Omega_{\phi}(dL_{z}/dt)$, where $\Omega_{\phi}$ is
the angular velocity on that orbit \cite{CKP}.  The quadrupole
formulae \erf{QuadEluminosity}--\erf{QuadLzluminosity} reproduce this
result for a circular orbit.  Since we use the correct input geodesic,
the semirelativistic approximation should and does give $\rmd E/\rmd
L_{z}$ accurately on the separatrix.  Since we have the correct result
in both extremes, it is perhaps less surprising that we also do quite
well at points in between.

\begin{figure}
%\centerline{\psfig{file=pePlot.ps,width=5cm,angle=0}}
\centerline{\includegraphics[keepaspectratio=true,height=5.in,
			     angle=0]{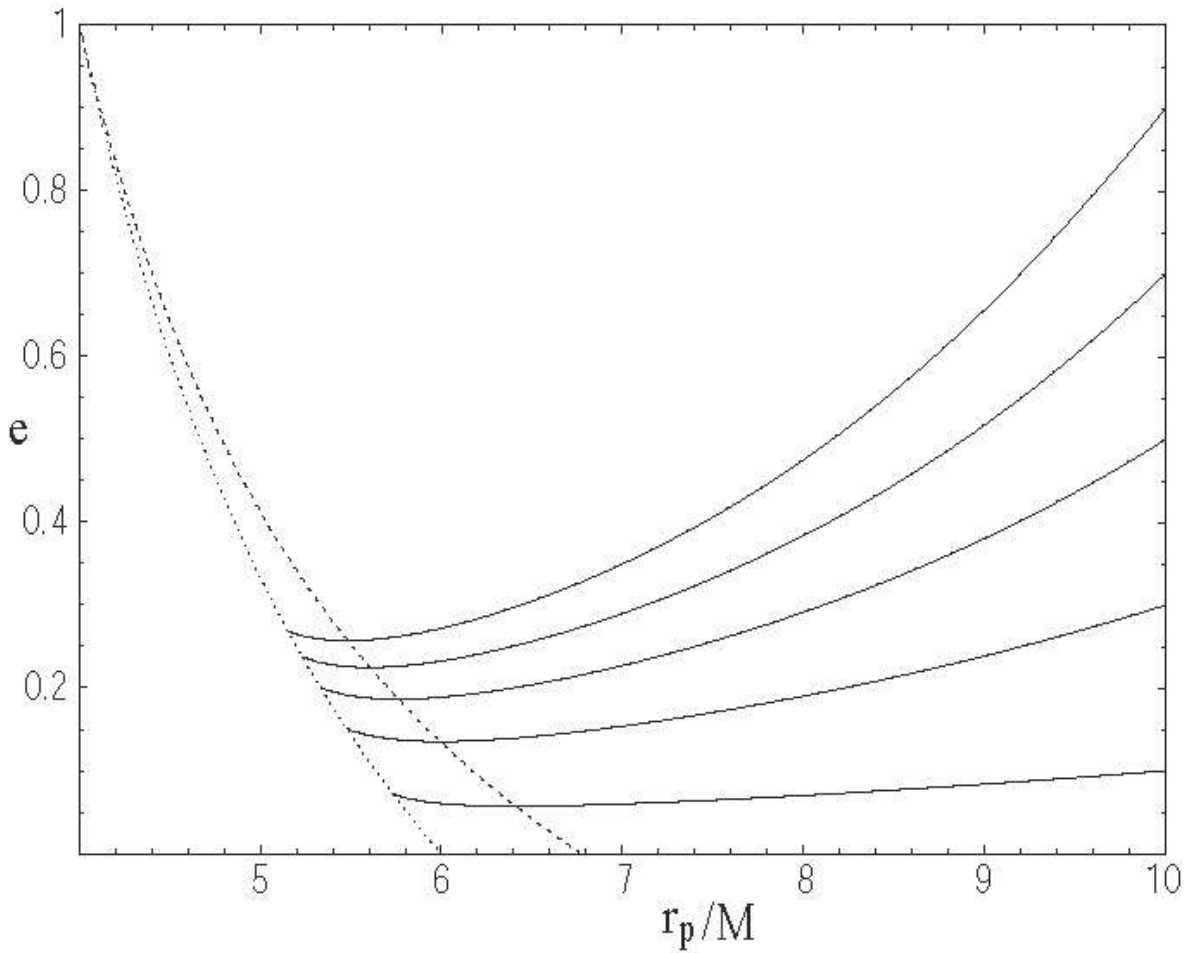}}
\caption{Sample inspiral trajectories in the $(r_{p}, e)$ plane.  The
solid lines illustrate inspiral trajectories.  The dotted line marks
the separatrix and all points to the left of this line are plunging
orbits.  The dashed line is the locus of points with $\rmd e/\rmd t =
0$.  In the region between this line and the separatrix, $\rmd e/\rmd
t > 0$ on all trajectories.}
\label{pefig}
\end{figure}

Figure~\ref{pefig} illustrates some inspiral trajectories in the
$(r_{p},e)$ plane.  The trajectory properties are determined largely
by two curves \--- the separatrix where the inspiral ends as the
particle plunges into the black hole and the locus $\rmd e/\rmd t =0$.
These are both marked on Figure~\ref{pefig}.  In general an inspiral
will begin with high eccentricity and moderate periapse.  Both the
periapse and eccentricity initially decrease, and this evolution
continues until the trajectory intersects the $\rmd e/\rmd t =0$
curve.  After that point, the periapse continues to decrease, but the
eccentricity increases until the trajectory reaches the separatrix and
the particle plunges.  As expected from previous arguments, these
general properties are in good agreement with results based on
Teukolsky calculations \cite{CKP} and quite different to Peters and
Mathews inspirals (which, for instance, have monotonically decreasing
eccentricity).  The location of the separatrix is a property of the
geodesics, and is therefore precisely reproduced in this
approximation.  The $\rmd e/\rmd t =0$ curve depends on the expression
used for the energy and angular momentum fluxes and is different here,
but only slightly.  In this approximation, the $\rmd e/\rmd t =0$
curve intersects the $e=0$ axis at $r_{p}=6.770\,M$, compared to
$r_{p}=6.681\,M$ in the Teukolsky case \cite{CKP}.

The increase in eccentricity prior to plunge is a generic feature of
EMRIs, but it is as much a property of the radial potential as it is
of the flux model.  As discussed earlier, realistic gravitational
waves will give
$\dot{E}=\Omega_{\phi}\,\dot{L_{z}}=((1+e)/(2\,(3+e)))^{\frac{3}{2}}\,
\dot{L_{z}}$ on the separatrix.  The leading order piece in both the
numerator and denominator of equation \erf{drpde} thus vanishes in the
limit $r_{p}\rightarrow r_{UCO}(e)$, but the leading correction to
both is from the logarithmic piece of $\rmd E/\rmd L_{z}$, and hence
we find $\rmd r_{p}/\rmd e < 0$.  However, this conclusion still holds
if the fluxes do not satisfy the circular orbit condition and the
cancellations {\it do not} occur.  The coordinate derivatives are such
that, independent of the value of $\rmd E/\rmd L_{z}$ on the
separatrix,
\begin{equation}
   \frac{\rmd r_{p}}{\rmd e} \approx 
-\frac{4\,(3+e)}{(1-e)\,(1+e)^{2}}
\label{drpdeExp}
\end{equation}
The nature of the potential thus forces either $r_{p}$ or $e$ to 
increase in the approach to plunge.

A final point to note from Figure~\ref{pefig} is that $\dot{e} \propto
e$ near $e=0$.  This property of the inspirals means that an initially
eccentric orbit cannot circularise in this model, although the
eccentricity at plunge can be arbitrarily small.  The property once
again derives from the circular condition
$\dot{E}=\Omega_{\phi}\,\dot{L_{z}}$, which ensures that circular
orbits remain circular under radiation reaction.  This is discussed in
more detail in \cite{GG05}, where corrections are given to enforce
this relation in kludged inspirals.  In the semirelativistic waveform
model, the condition is automatically satisfied and no correction is
required.

\subsection{``Kludge'' waveform inconsistency}\label{KludgeInconsis}
\begin{figure}
\centerline{\includegraphics[keepaspectratio=true,height=4.in,
			     angle=-0]{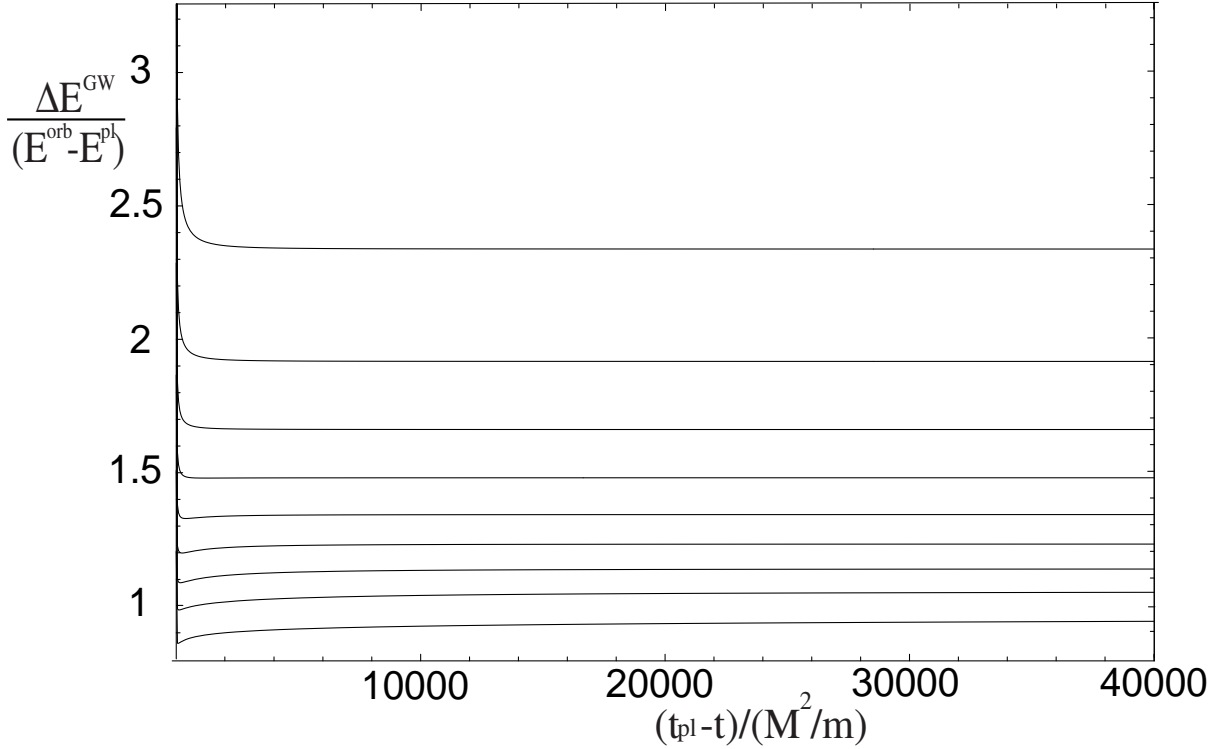}}
\caption{Ratio of the energy content of ``kludge'' gravitational waves
to the change in energy of the source particle orbit, relative to the
energy at plunge.  This is shown as a function of time until plunge
(in unite of $M^2/m$) for eccentricities at plunge $e_{pl}=0.1$,
$0.2$...$0.9$ (from lowermost curve to uppermost).}
\label{KludgeError}
\end{figure}

As mentioned in the introduction, waveforms based on the
semirelativistic approximation are being used extensively to scope
out LISA data analysis \cite{GHK02,GG05,tev05}.  The waveforms are
generated by first constructing an inspiral trajectory and then using
the semirelativistic construction of the quadrupole moment tensor to
compute a waveform.  In the most basic form of this ``kludge''
\cite{GHK02}, the phase space trajectory for inspirals into
non-spinning black holes is computed by integrating the Peters and
Mathews energy and angular momentum fluxes \erf{PMdEdt}, \erf{PMdLdt}.
This leads to an inconsistency since the energy and angular momentum
content of the gravitational waves differs from the energy and angular
momentum lost by the inspiralling particle that is nominally generating
the radiation.  We can estimate this inconsistency using the
semirelativistic results.  A phase space trajectory is generated
using the fluxes \erf{PMdEdt}, \erf{PMdLdt} and equation \erf{drpde}.
We choose to specify the eccentricity of the inspiral at plunge and
integrate backwards along the inspiral trajectory.  By integrating the
semirelativistic fluxes along this trajectory, we calculate the total
energy and angular momentum content of the gravitational waves.
Figure~\ref{KludgeError} shows the ratio of the gravitational wave
energy flux to the change in energy of the particle orbit as a
function of the time {\it until plunge} (in units of $M^2/m$).  Time
along the inspiral trajectory therefore decreases toward the right.
There is a curve for each eccentricity at plunge from $e=0.1$ to
$e=0.9$ in intervals of $0.1$.  We see that there is a significant
inconsistency in the kludge waveforms.  For low eccentricity at
plunge, the kludge gravitational waves contain less energy than they
should, but for eccentricity at plunge greater than about $0.25$, they
contain too much energy, as much as a factor of three in extreme
cases.  This means that signal-to-noise ratios (SNRs) computed from these
waveforms are likely to be overestimates of the true SNRs.  It is not
clear from these results whether this discrepancy will be larger or
smaller when the central black hole is spinning, but this will be
investigated in the future \cite{tev05}.  However, it is important to
be aware of the existence and magnitude of this problem when
interpreting results based on the kludge waveforms.  If
semirelativistic fluxes were used to integrate the phase space
trajectories, there would be no such inconsistency and this might
therefore be another future application of these results, once they
are extended to spinning black holes.

\subsection{Analytic results}\label{sec:Analytic}
The previous results have shown the usefulness of the
semirelativistic approximation, but the described method, based on
integration of the geodesic equations, is not easy to implement in
numerical simulations. In this section, we present some analytic
results based on the semirelativistic approximation which can be
easily evaluated without numerical integration of orbits.

\subsubsection{Fitting functions for $\Delta E$ and $\Delta
L_{z}$}\label{fitfunc} A useful tool for simulations is a fitting
function that has a simple form and which reproduces the
semirelativistic results for $\Delta E$ and $\Delta L_{z}$ to
reasonable accuracy.  For a geodesic of given eccentricity, the
periapse can have any value between the UCO for that eccentricity
\erf{rUCO} and infinity.  For large values of the periapse, the orbit
is entirely within the weak-field region of the spacetime.  The orbit
and radiation will therefore look very like those from a Keplerian
orbit, as described by expressions \erf{PMdEdt}--\erf{PMdLdt}
\cite{PM63}.  Multiplying these expressions by the Keplerian orbital
period, the energy and specific angular momentum lost on a single pass
may be found to be
\begin{eqnarray}
   \Delta E &=& -\frac{64\,\pi}{5} \, \frac{m}{M} \,
   \frac{1}{\left(1+e\right)^{\frac{7}{2}}}\,\left(1+\frac{73}{24}\,
   e^{2}+\frac{37}{96}\,e^{4}\right)\,
   \left(\frac{r_{p}}{M}\right)^{-\frac{7}{2}} \label{PMdE} \\
   \Delta L_{z} &=& -\frac{64\,\pi}{5}\,m\,
   \frac{1}{\left(1+e\right)^{2}}\,
   \left(1+\frac{7}{8}\,e^{2}\right)\,\,
   \left(\frac{r_{p}}{M}\right)^{-2} \,.  \label{PMdL}
\end{eqnarray}
In the parabolic case ($e=1$), these become
\begin{eqnarray}
   \Delta E &=& -\frac{85\,\pi}{12\,\sqrt{2}} \, \frac{m}{M} \,
   \left(\frac{r_{p}}{M}\right)^{-\frac{7}{2}} \label{PMpardE} \\
   \Delta L_{z} &=& -6\,\pi\,m\,\left(\frac{r_{p}}{M}\right)^{-2} \,.
   \label{PMpardL}
\end{eqnarray}
As the periapse approaches the UCO, the energy and angular momentum
lost per pass increases.  In fact, ignoring radiation reaction, the
energy and angular momentum losses diverge as $r_{p} \rightarrow
r_{UCO}$.  This is because the geodesic with $r_{p}=r_{UCO}$ spends
an infinite amount of time whirling around the black hole with
$r\approx r_{UCO}$.  In practice, radiation reaction will prevent
this situation arising (for a discussion of the transition from
inspiral to plunging orbits see \cite{ROS03}).  However, the
energy and angular momentum lost should increase rapidly as the
periapse approaches $r_{UCO}$, since the orbit whirls around the
black hole an increasing number of times.

As discussed previously, the whirl behaviour is a property of the
geodesics.  Thus, although the semirelativistic treatment of the
radiation is only approximate, one will still see this divergent
behaviour as $r_{p} \rightarrow r_{UCO}$, since the source for the
radiation is an exact geodesic trajectory.  This feature is missing in
the Keplerian treatment \cite{PM63}.  For an extreme zoom and whirl
orbit, most of the gravitational radiation is emitted during the whirl
phase, when the particle is on an approximately circular orbit.  It is
reasonable to guess that the total losses due to gravitational
radiation are roughly proportional to the number of times the particle
whirls around the black hole.

In the parabolic case, one can estimate the number of whirls by
computing the proper time taken for the orbit to pass from periapse to
some ``whirling radius'', $r_{w}$.  This is found using \erf{rdotrpe}
to be
\begin{equation}
   \tau = \frac{1}{\sqrt{2\,M}} \int_{r_{p}}^{r_{w}}
   \frac{r^{\frac{3}{2}}}{\sqrt{\left(r-r_{p}\right)
   \left(r-\frac{2\,M\,r_{p}}{ r_{p}-2\,M}\right)}} \, \rmd r
   \label{tauwhirl} \,.
\end{equation}
The radius does not change significantly during the whirl phase, so
approximating the numerator by $r_{p}^{3/2}$
\begin{equation}
\tau \approx \frac{1}{\sqrt{2\,M}} r_{p}^{\frac{3}{2}} \cosh^{-1}
\left(\frac{2(r_{p}-2\,M)}{r_{p}(r_{p}-4\,M)}
\left[r_{w}-\frac{r_{p}^{2}} {2(r_{p}-2\,M)}\right] \right)
\label{tauwhirl2} \,.
\end{equation}
Using the same assumption, $\rmd \phi/\rmd \tau \approx L/r_{p}^{2}$
and we estimate that while $r < r_{w}$, the number of azimuthal
cycles that the particle completes is
\begin{equation}
\label{nwhirls} N_{whirls} =\frac{\Delta \phi}{2\pi} \approx
\sqrt{\frac{r_{p}}{r_{p}-2\,M}} \frac{1}{\pi} \cosh^{-1}
\left(\frac{2(r_{p}-2\,M)}{r_{p}(r_{p}-4\,M)} \left[
r_{w}-\frac{r_{p}^{2}}{2(r_{p}-2\,M)}\right] \right) \,.
\end{equation}
The radius $r_{w}$ should be chosen to define the start and end of the
whirl phase.  Our objective is to guess a functional form that
approximates the energy and angular momentum loss when $r_{p}\approx
r_{UCO}$ and we assume that $\rmd E$ and $\rmd L_{z}$ are proportional
to \erf{nwhirls} in that limit.  This is likely to be a particularly
good approximation for highly eccentric orbits, in which the `zoom'
and `whirl' phases are quite distinct.  Appropriate fitting functions
should approach \erf{PMdE} and \erf{PMdL} in the limit $r_{p}
\rightarrow \infty$ and should diverge like \erf{nwhirls} as
$r_{p}\rightarrow r_{UCO}$.  Working once again in the parabolic case,
the simplest such function is
\begin{equation}
   \frac{M}{m}\,\Delta E = A^{E}\cosh^{-1}\left[1+B^{E}\,
   \left(\frac{4\,M}{r_{p}}\right)^{6}
   \,\frac{M}{r_{p}-4\,M}\right] + C^{E} \,
   \left(\frac{r_{p}}{M}-4\right)\,\left(\frac{M}{r_{p}}
   \right)^{\frac{9}{2}}\,.
   \label{dEfitform}
\end{equation}
One could fix all three coefficients by matching the behaviour in the
limit $r_{p}\rightarrow r_{UCO}$, but \erf{dEfitform} will not then
necessarily reproduce the asymptotic form \erf{PMpardE}.  Instead, we
fix
$A^{E}$ and $B^{E}$ using an expansion near $r_{p}=r_{UCO}$ and then
fix $C^{E}=-(85\pi/(12\sqrt{2}) + 64A^{E}\sqrt{2B^{E}})$ to match
\erf{PMpardE} asymptotically.

\begin{figure}
\centerline{\includegraphics[keepaspectratio=true,height=4.in,
			     angle=0]{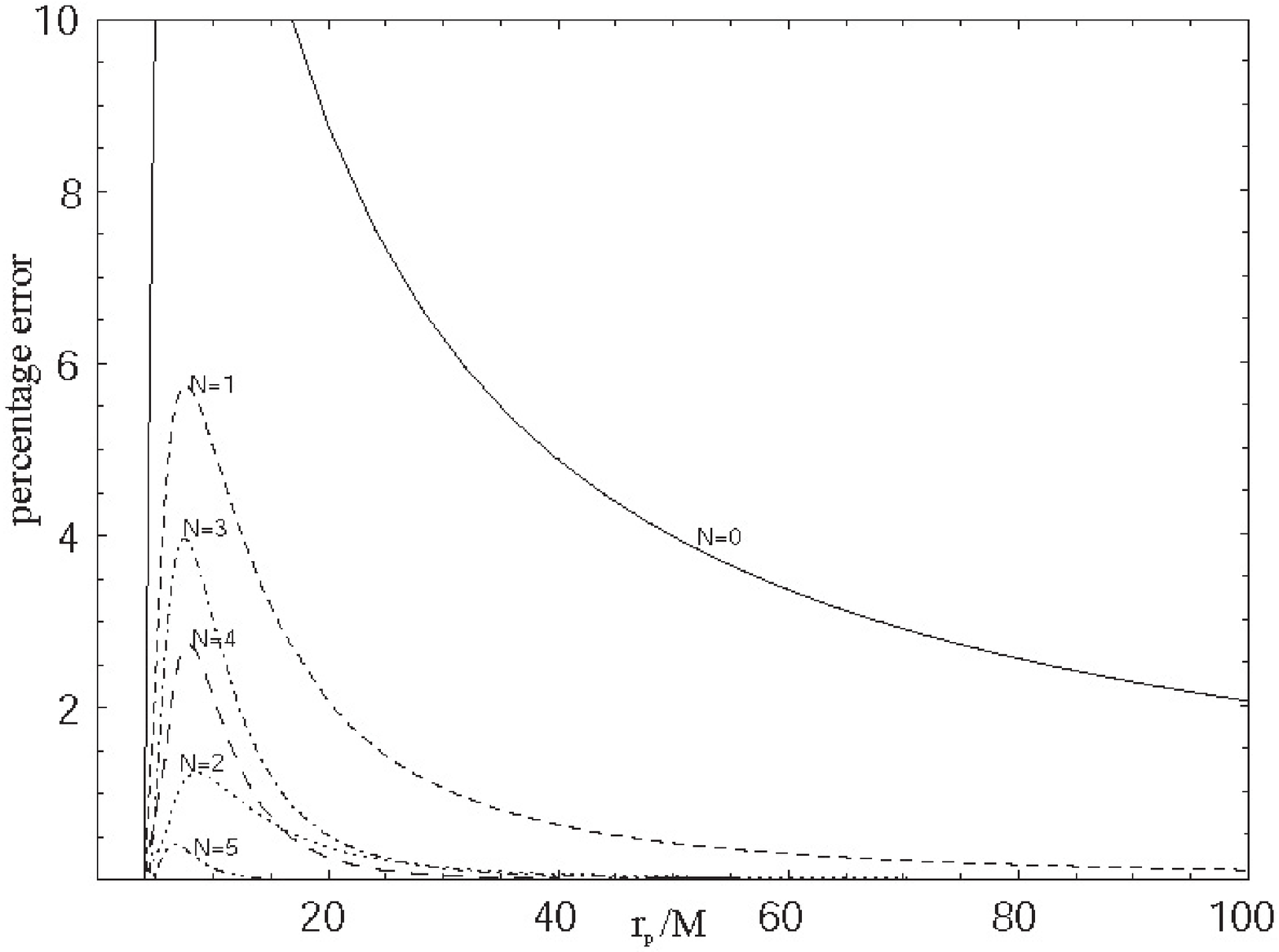}}

\vspace{0.2in}
\centerline{\includegraphics[keepaspectratio=true,height=4.in,
			     angle=0]{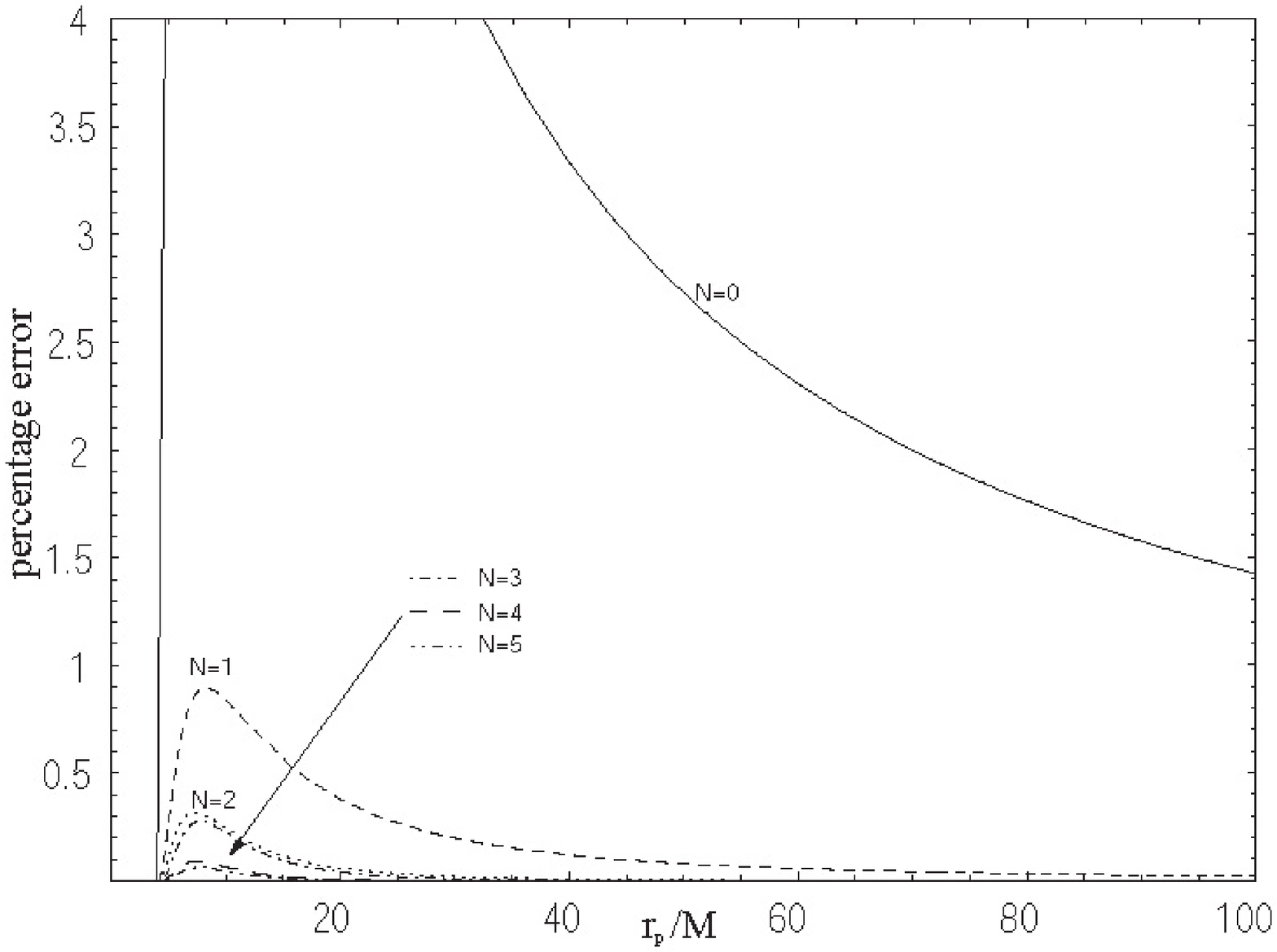}}

\caption{Error using fitting functions to approximate the analytic
expressions for the energy (upper plot) and angular momentum loss
(lower plot).  In each plot, the absolute percentage error in the fit
is shown for fitting functions of various orders, $N=2,\cdots,6$.}
\label{fitconvfig}
\end{figure}

The next section will demonstrate how exact expressions for the
energy and angular momentum radiated in our model may be obtained in
terms of elliptic integrals.  Using these full analytic expressions,
we can predict the values of the fitting function coefficients
\begin{equation}
   A^{E}=-\frac{\sqrt{2}}{10}=-0.141421, \qquad B^{E}=0.752091,\qquad
   C^{E}=-4.634643 \,.  \label{dEfitcoeffs}
\end{equation}
The equivalent fitting function for the angular momentum lost is
\begin{equation}
   \frac{\Delta L_{z}}{m} = A^{L_{z}}\cosh^{-1}\left[1+B^{L_{z}}\,
   \left(\frac{4\,M}{r_{p}}\right)^{3}\,\frac{M}{r_{p}-4\,
   M}\right]+ C^{L_{z}} \,\left(\frac{r_{p}}{M}-4\right)\,
   \left(\frac{M}{r_{p}}\right)^{3}
   \label{dLfitform}
\end{equation}
with coefficients
\begin{equation}
   A^{L_{z}} =-\frac{4\sqrt{2}}{5}=-1.13137,
   \qquad B^{L_{z}}=1.31899, \qquad
   C^{L_{z}}=-\left(6\,\pi+8\,A^{L_{z}}\,\sqrt{2\,
   B^{L_{z}}}\right)=-4.149103\,.
   \label{dLfitcoeffs}
\end{equation}
The fitting function \erf{dEfitform} can be used to match the lowest
order terms in an expansion of $\Delta E$ near $r_{p}=r_{UCO}$ and
$r_{p} \rightarrow \infty$.  It is possible to add additional terms to
give a more general function which can match $\Delta E$ at arbitrary
orders
\begin{eqnarray}
  \nonumber \fl \frac{M}{m}\,\Delta E &=& \left(\sum_{n=0}^{N}
  A^{E}_{n}\,\left(\frac{M\,(r_{p}-4\,M)}{r_{p}^{2}}\right)^{n}
  \right)\,\cosh^{-1}\left[1+B^{E}_{0} \,
  \left(\frac{4\,M}{r_{p}}\right)^{N_{E}-1}\,
  \frac{M}{r_{p}-4\,M}\right] \\ & & +
  \frac{M^{\frac{N_{E}}{2}}\,(r_{p}-4\,M)}{r_{p}^{1+
  \frac{N_{E}}{2}}} \,\sum_{n=0}^{N}C^{E}_{n}\,\left(\frac{M\,
  (r_{p}-4\,M)}{r_{p}^{2}}\right)^{n} +
  \frac{M^{1+\frac{N_{E}}{2}}\,(r_{p}-4\,M)}{r_{p}^{2+
  \frac{N_{E}}{2}}}\, \sum_{n=0}^{N-1}
  B^{E}_{n+1}\,\left(\frac{M\,(r_{p}-4\,M)}{r_{p}^{2}}\right)^{n}
\label{genfitform}
\end{eqnarray}
In this, we fix $N_{E}=7$ to give the correct leading order behaviour
\erf{PMpardE} as $r_{p}\rightarrow\infty$.  The parameter $N$
indicates the order of the fit, i.e., the number of terms we include.
The second and third series have terms in common, but writing the
expansion in this way allows one to read off consecutive coefficients
easily.  The next section will show that expanding $\Delta E$ about
the separatrix gives terms in $(r_{p}-4M)^{j}\,\ln (r_{p}-4M)$ and in
$(r_{p}-4M)^{k}$, while an expansion as $r_{p}\rightarrow\infty$ gives
terms of the form $1/r_{p}^{\frac{N_{E}}{2}+l}$.  The coefficient of
the $j=0$ term gives $A_{0}^{E}$, then the $k=0$ term gives
$B_{0}^{E}$ and the $l=0$ term gives $C^{E}_{0}$.  Continuing in this
way, the $j,k,l=n$ terms determine $A_{n}^{E}$, $B_{n}^{E}$ and
$C_{n}^{E}$ respectively.  Thus, an expansion to order $N$ will match
the lowest $N+1$ terms in $j$, $k$ and $l$.  A similar fitting form
may be used for $\Delta L_{z}/M$, but with $N_{E}$ replaced by
$N_{L_{z}}=4$ (once again, to reproduce the correct leading order
behaviour \erf{PMpardL} as $r_{p} \rightarrow \infty$).
Figure~\ref{fitconvfig} illustrates how the fitting functions converge
as the order of the fit, $N$, is increased.  We see that as $N$
increases, the fit improves at large radii, but initially gets
slightly worse at moderate radii before converging there also.  The
$N=2$ fit is accurate to about one percent everywhere, so we include
these parameters here
\begin{eqnarray}
\nonumber A^{E}_{0}&=&-0.141421, \qquad A^{E}_{1}=0., \qquad 
A^{E}_{2}=-1.20797, \qquad B_{0}^{E} =0.752091, \qquad 
B^{E}_{1}=-103.215, \\ \nonumber B^{E}_{2}&=&727.515, \qquad 
C^{E}_{0}=-4.63464, \qquad C^{E}_{1}=69.1683, \qquad C^{E}_{2} 
=-439.378 \\ \nonumber A^{L_{z}}_{0}&=&-1.13137, \qquad 
A^{L_{z}}_{1}=0., \qquad A^{L_{z}}_{2}=0., \qquad 
B_{0}^{L_{z}}=1.31899, \qquad B^{L_{z}}_{1}=-53.4491, \\ 
B^{L_{z}}_{2}&=&29.7857, \qquad C^{L_{z}}_{0}=-4.1491, \qquad 
C^{L_{z}}_{1}=25.4129, \qquad C^{L_{z}}_{2}= 15.1726
\label{PSGfitcoeffs}
\end{eqnarray}

This fitting function was derived using simple arguments about how the
energy and angular momentum lost behave.  These arguments are valid
in general for radiation that is produced by a body orbiting in the
Schwarzschild potential and will apply to fluxes computed using the
Teukolsky formalism. In a separate paper \cite{GKL1} we derive a fit
of this form to Teukolsky data computed for parabolic orbits in
\cite{martel04}, which even for $N=2$ is accurate to $<0.2\%
$ everywhere.

This simple fitting function is clearly a useful and accurate way to
evolve EMRI orbits.  In the case of arbitrary eccentricity, a similar
type of fitting function can be derived, but the coefficients
$A_{0}^{E}$ etc.  are now functions of eccentricity.  In the
semirelativistic approximation, the functions can be evaluated
explicitly.  This is discussed in more detail in
appendix~\ref{fitapp}.  It is more complicated to compute a fit to
Teukolsky-based fluxes, since the coefficients in the expansion must
be further expanded as functions of eccentricity.  However, it should
be possible to derive a reasonable fit using a polynomial ansatz, of
the form suggested by the semirelativistic results.  Once sufficient
Teukolsky-based data is available, this fitting procedure will allow
us to generate a comparatively simple analytic expression for use in
computation of EMRIs.

\subsubsection{Exact expression}\label{sec:analexp}
As mentioned in the preceding section, it is possible to derive exact
analytic expressions for the radiation loss predicted by our
quadrupole approximation.  This is possible because in any
axisymmetric spacetime, the rate of energy loss at a given moment in
time cannot depend on the absolute value of the $\phi$ coordinate of
the particle, since a shift $\phi \rightarrow \phi+\phi_{0}$ will
leave the spacetime unchanged (note that the energy flux {\em in a
given direction} will be dependent on the relative difference in
$\phi$ between the source and the observer).  In the quadrupole
approximation used here to compute the gravitational radiation,
$dE/dt$ is given by the square of the third time derivative of the
quadrupole moment tensor.  This is a function only of the particle
coordinates $r(t)$ and $\phi(t)$.  By the axisymmetry argument, the
expression for $dE/dt$ can depend only on $r$, $\dot{r}$, $\ddot{r}$,
$\dddot{r}$, $\dot{\phi}$, $\ddot{\phi}$ and $\dddot{\phi}$.  For
geodesics of the Schwarzschild spacetime (also for equatorial orbits
in the Kerr spacetime), $\rmd\phi/\rmd\tau$ and $\rmd t/\rmd\tau$ are
rational functions of $r$ only and $(\rmd r/\rmd \tau)^{2}=V(r)$ is a
cubic or quartic polynomial in $r$.  Any derived expression, such as
$\rmd E/\rmd r =(\rmd E/\rmd t)/(\rmd r/\rmd t)$, will therefore be a
rational function of polynomials in $r$ and $\sqrt{V(r)}$.  It is
known \cite{abram64} that the integral of any rational function of
polynomials in $x$ and $y$, where $y^{2}$ is a cubic or quartic
polynomial in $x$, can be expressed in terms of elliptic integrals.
One can therefore write the energy and angular momentum radiated in
closed form in terms of elliptic integrals.

By substitution of the geodesic equations \erf{phidot}, \erf{tdot} and
\erf{rdotrpe} into \erf{QuadEluminosity}--\erf{reducedQM}, we may
write $\rmd E/\rmd t$ and $\rmd L_{z}/\rmd t$ as functions of $r$ and
then integrate over one orbit.  In the parabolic case, the energy loss
is found to be
\begin{eqnarray}
  \frac{M}{m}\,\Delta E = &-&\frac{8\,{\sqrt{2}}\,M^{\frac{21}{2}}}
  {{1673196525\,{\left(r_{p}
  -2\,M \right) }^2\, {r_{p}}^{\frac{17}{2}}}}  \left[{\bf
  E}\left(\sqrt{\frac{2\,M}{r_{p}-2\,M}} \right)
  f_{1}\left(\frac{r_{p}}{M}\right) + {\bf
  K}\left(\sqrt{\frac{2\,M}{r_{p}-2\,M}} \right)
  f_{1}\left(\frac{r_{p}}{M}\right)\right]
  \label{PardeltaE}
\end{eqnarray}
where
\begin{eqnarray}
  f_{1}(y)= &-&2y\,\left( 27850061568 - 83550184704\,y +
  117662445984\,{y}^2 - 102686941680\,{y}^3  \right. \nonumber \\ 
  && \left.+ 64808064704\,{y}^4 - 33026468872\,{y}^5 +
  12784148218\,{y}^6 - 2873196259\,{y}^7  \right.  \nonumber \\ 
  && \left.  + 185808888\,{y}^8 + 17119626\,{y}^9 +
  2451526\,{y}^{10} +368640\,{y}^{11} + 20480\,{y}^{12} \right)
  \nonumber
  \end{eqnarray}
and
\begin{eqnarray}
    f_{2}(y) = &&\left( -72901570560 + 274404834816\,y
    - 424693524096\,{y}^2  \right. \nonumber \\ 
  && \left.+ 378109481088\,{y}^3 - 249480499840\,{y}^4 +
  154011967968\,{y}^5  \right. \nonumber \\ 
  && \left.  -84437171728\,{y}^6 + 31689370996\,{y}^7 -
  6231594434\,{y}^8 + 321950817\,{y}^9  \right. \nonumber \\ 
  && \left.  + 27462280\,{y}^{10} + 4073612\,{y}^{11} +
  696320\,{y}^{12} + 40960\,{y}^{13} \right)\ . \nonumber 
\end{eqnarray}
In this, ${\bf K}$ and ${\bf E}$ are the complete Elliptic integrals
of the first and second kinds respectively, defined by
\cite{grad94,abram64}
\begin{equation}
  {\bf K}(k) = \int_{0}^{\frac{\pi}{2}} \,\, \frac{\rmd
  \phi}{\sqrt{1-k^{2}\,\sin^{2}\phi}}, \qquad {\bf E}(k) =
  \int_{0}^{\frac{\pi}{2}} \,\,\sqrt{1-k^{2}\,\sin^{2}\phi} \,\, \rmd
  \phi.  \label{ellints}
\end{equation}
The corresponding result for the angular momentum lost is
\begin{equation}
  \frac{\Delta L_{z}}{m} = \frac{64\,M^7}{
  24249225\,{r_{p}}^{\frac{11}{2}}\left(r_{p}-2\,M \right)^
  {\frac{3}{2}}}\left[ {\bf E}\left(\sqrt{\frac{2\,M}{r_{p}-2\,M}}
  \right) g_{1}\left(\frac{r_{p}}{M}\right) + {\bf
  K}\left(\sqrt{\frac{2\,M}{r_{p}-2\,M}} \right)
  g_{2}\left(\frac{r_{p}}{M}\right) \right]
  \label{PardeltaLz}
\end{equation}
where
\begin{eqnarray}
  g_{1}(y) = && y \left( 181817664 - 363635328\,y + 245236248\,{y}^2 -
  49673460\,{y}^3  \right.  \nonumber \\
  && \left.  - 7833906\,{y}^4 + 2016105\,{y}^5 + 283252\,{y}^6 +
  35896\,{y}^7 + 4120\,{y}^8 \right) \nonumber
\end{eqnarray}
and
\begin{eqnarray}
   g_{2}(y) = &&\left( 71285760 - 324389184\,y -
   468548880\,{y}^2 - 277856496\,{y}^3
  + 54521424\,{y}^4 \right.  \nonumber \\
  && \left.  + 6181872\,{y}^5 - 1630457\,{y}^6 -
  238086\,{y}^7 - 31776\,{y}^8 - 4120\,{y}^9 \right) \nonumber
\end{eqnarray}
These exact expressions can be used to derive the fitting function
described in the previous section.  As $r_{p} \rightarrow \infty$, the
argument of the elliptic integrals, $\sqrt{2\,M/(r_{p}-2\,M)}
\rightarrow 0$.  In a series expansion of the integrals about $k=0$
the lowest five orders in $k$ cancel and one successfully recovers
\erf{PMpardE}--\erf{PMpardL}.

As $r\rightarrow r_{UCO}=4\,M$, the argument of the elliptic
integrals $\sqrt{2\,M/(r_{p}-2\,M)} \rightarrow 1$ and the
elliptic integrals diverge.  Using \cite{abram64} and some algebraic
manipulation, the asymptotic forms of the elliptic integrals as $k
\rightarrow 1$ are found to be
\begin{eqnarray}
   {\bf K}(k) &=& -\frac{1}{2}\,\ln{\left(1-k^{2}\right)}\,\,
   +2\,\ln{2}\,\,-\frac{1}{8}\,\left(1-k^{2}\right)\,
   \ln{\left(1-k^{2}\right)}\,\,+O\left(1-k^{2}\right)
   \label{asymK} \\
   {\bf E}(k) &=& 1\,\,-\frac{1}{4}\,\left(1-k^{2}\right)\,
   \ln{\left(1-k^{2}\right)}\,\,+\left(\ln{2}\,\,
   -\frac{1}{4}\right)\,\left(1-k^{2}\right)\,\,
   +O\left((1-k^{2})^{2}\,ln{\left(1-k^{2}\right)}\right).
   \label{asymE}
\end{eqnarray}
The asymptotic form of \erf{PardeltaE}--\erf{PardeltaLz} as
$r_{p}\rightarrow 4\,M$ is
\begin{equation} \label{asymth}
   \frac{\rmd X}{\eta} \approx
   p_{X}\,\ln{\left(\frac{r_{p}}{M}-4\right)}\,+\,q_{X}\,+O\left(
   \frac{r_{p}}{M}-4\right)
\end{equation}
In this, $X$ refers to either $E$ or $L_{z}/M$.  The values of the
constants are
\begin{eqnarray}
   p_{E} &=& \frac{1}{5\,\sqrt{2}}, \qquad q_{E} = 2\,\left(
   \frac{16370483137}{53542288800\,{\sqrt{2}}} -\frac{\ln
   (2)}{2\,{\sqrt{2}}} \right), \\ p_{L_{z}}&=&\frac{4\sqrt{2}}{5},
   \qquad q_{L_{z}} = 2\,\left( \frac{1613849\,{\sqrt{2}}}{1616615} -
   2\,{\sqrt{2}}\,\ln (2) \right).
\end{eqnarray}
The fitting functions \erf{dEfitform}--\erf{dLfitform} may be
similarly expanded near $r_{p} = r_{UCO} = 4\,M$
\begin{equation}
   \frac{M}{m}\,\Delta X = -A_{X}\ln{\left(\frac{r_{p}}{M}-
   4\right)}+A_{X}\ln{(2b_{X})}+O\left(\frac{r_{p}}{M}-4\right)
   \label{fitasym}
\end{equation}
In this, $X$ once again refers to either $E$ or $L_{z}/M$, and
$N_{E}=7$, $N_{L_{z}}=4$.  Equating \erf{asymth} and \erf{fitasym},
one obtains the coefficients of the fitting function given earlier
\erf{dEfitcoeffs}, \erf{dLfitcoeffs}.

A similar analysis can be performed for orbits of arbitrary
eccentricity, and is described in appendix~\ref{exactapp}.  The exact
expressions are somewhat cumbersome and we recommend using the fitting
function in most applications, since this performs extremely well.
The exact expressions have been included for completeness, and to help
explain why the fitting function works.

\subsection{Hyperbolic captures}
\begin{figure}
\centerline{\includegraphics[keepaspectratio=true,height=6.in,
angle=0]{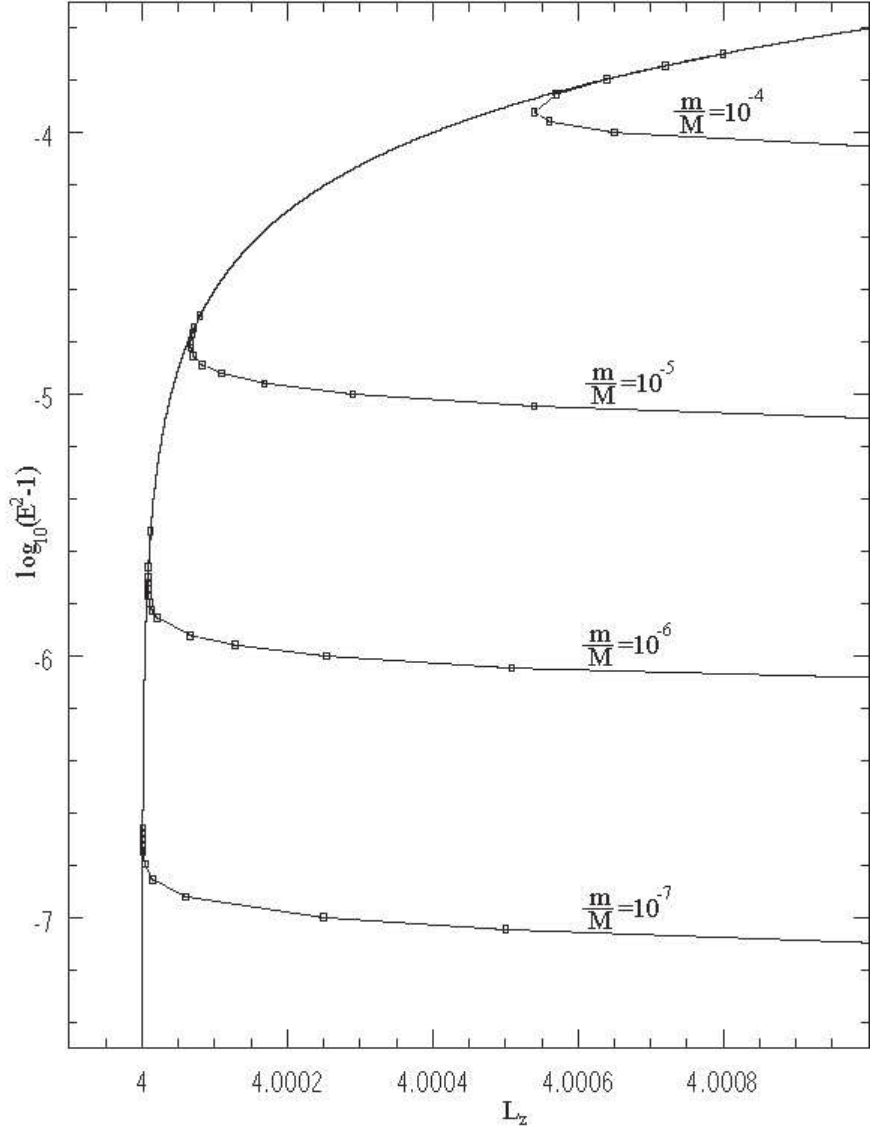}}
   \caption{``Hyperbolic'' orbits (i.e., orbits with $E > 1$) that are
   captured after one close encounter with the black hole, for various
   mass ratios.  Orbits whose energy and angular momentum place them
   above and to the right of the line for a given mass ratio remain
   unbound and are not captured.  The line which begins at bottom left
   and curves around to top right in the figure is the separatrix line
   separating unstable plunging orbits (to its left) from stable
   orbits.}
\label{HypCapt}
\end{figure}

In this paper we mostly focus on parabolic orbits, which serve as a
useful model for all orbits which are likely to lead to sources of
interest to LISA, since such orbits will always initially have
eccentricities very close to one.  However, objects can also be
captured from orbits with $e >1$.  In such cases the orbit is unbound,
but may ultimately inspiral if it makes a close approach to the
central black hole and loses enough energy and angular momentum in
doing so to become bound.  Our results suggest that if the angular
momentum of this orbit is low (close to the minimum $L_{z}=4$), then
the scattered body will become bound if it is on an orbit whose energy
$E$ is such that $E^2-1 < m/M$ roughly speaking, where $m/M$, the mass
ratio, is small.  For larger angular momenta the amount of excess
energy which can be radiated away on the first pass is smaller, and so
the orbital energy must be even closer to unity for the body to become
bound.  Figure~\ref{HypCapt} shows which hyperbolic orbits can lead to
captures, for low orbital angular momenta. The energy and angular momentum lost
to gravitational waves by a particle on a hyperbolic orbit are given by the same
equations \erf{GendeltaE}--\erf{GendeltaL} that apply to bound orbits, just by
inserting $e>1$ consistent with the definitions \erf{Eofrpe}--\erf{Lofrpe}\footnote{This statement, as appeared in the published version of this paper, is not correct. The published erratum has been appended in Appendix~\ref{erratum}.}.
Figure~\ref{HypCapt} was generated by using equations
\erf{Eofrpe}--\erf{Lofrpe} in conjunction with equation \erf{GendeltaE} to write $\Delta E = (m/M) F_E(E, L_z)$ for hyperbolic orbits. Points on the curves obey the equation
\begin{equation}
E - 1 = -\frac{m}{M} \, F_{E} ( E, L_z). \label{hypgraphEq}
\end{equation}
Fixing the energy, the angular momentum solution to \erf{hypgraphEq} is obtained by iteration.

This figure indicates that there is another type of orbit which
becomes bound after the first pass \-- those that are close to the
separatrix.  If the orbit is very close to the plunge line, it will
also lose enough energy to become bound even if it has much more
energy than $E^2-1=m/M$.  The reason seems obvious after a glance at
Figure~\ref{ZWpot}.  If the energy of the scattered body is
sufficiently close to 1 then it is close enough to the potential well
in which bound orbits exist to lose sufficient energy on one pass to
fall into the well.  If the energy and angular momentum of the orbit
are such that the particle's energy places it at the maximum of the
potential, then the particle ``whirls'' around the central black hole
at the radius of the potential maximum.  The scattered body thus
spends an abnormally long time near periastron and hence radiates an
unusually large amount of energy, enough to become bound.  Of course
these bodies will generally plunge rather soon after capture because
the amount of angular momentum radiated will also decrease the height
of the potential barrier.  In fact many of these orbits will plunge on
their first pass, having dissipated enough angular momentum to shrink
the potential barrier so they pass over it into the plunging region
beyond.  One has to keep in mind that our adiabatic approximation
breaks down as one approaches the line separating stable from plunging
orbits (depending on the mass ratio), so that we can give no
definitive picture of what occurs in this regime except to say that
the general behaviour is probably correct.  Readers interested in the
transition from inspiral to plunge should consult \cite{ROS03}.

Orbits that are scattered close to the separatrix line are
``captured'', in the sense that they either plunge or become bound.
Particles which are close to being parabolic orbits are also captured
and may serve to modestly increase the capture rate for LISA.
Particles passing very near to the black hole must thus pass between
Scylla and Charybdis \footnote{In {\it The Odyssey} Scylla and
Charybdis were monsters who guarded opposite sides of a narrow
straight through which ships must pass on their way from Troy in Asia
Minor to Ithaca in Greece.  Scylla was a six-headed reptile who
plucked sailors from the decks of their ships as they passed, while
Charybdis was an undersea monster which created a great whirlpool by
its sucking in of seawater, the whirlpool causing ships to be pulled
under if they strayed too near.}.  If they pass too close to Scylla,
through having energy only marginally greater than 1, then they are
plucked from their unbound orbit by gravitational radiation reaction
and end up in a bound orbit.  If they approach with too great an
energy for their angular momentum then they are sucked down by
Charybdis and plunge into the black hole itself.

\section{Results and discussion}\label{sec:Discussion}
As with earth-based gravitational wave detectors like the Laser Interferometer Ground Observatory (LIGO), theoretical predictions of source event rates and signal
characteristics for LISA will play an important role in the successful
operation of the observatory.  Up to now, Newtonian-order estimates
(like Peters and Mathews) have been widely relied upon to estimate
waveforms and fluxes from extreme-mass ratio inspirals, even though
much of what is of interest to LISA, even in the early stages of
inspiral, occurs inside the region of relatively strong curvature
close to the central black hole.  The principle reason for this is
simply ease of use.  Even when accurate methods, such as the Teukolsky
formalism and self-force calculations, prove capable of dealing with
arbitrary orbits they may still be slow and cumbersome for many
applications.  This paper attempts to make available a range of
techniques which combine ease of use with fairly robust accuracy over
almost the whole inspiral of an extreme-mass-ratio binary. These
results are of particular use for highly eccentric orbits, where
frequency domain Teukolsky calculations perform poorly \cite{GK02}
and time domain codes have not yet been fully developed
\cite{martel04}.

The key elements to take away from this study of the
semirelativistic approximation are:

\begin{itemize}

\item Simple analytic expressions to estimate the fluxes $\Delta E$
and $\Delta L_{z}$, suitable for use in computational endeavours.

\item The optimal choice of parameters with which to describe orbits
which stray near the central black hole are the geodesic parameters
$\{r_{p},e\}$ rather than $\{E, L_{z}\}$: the waveforms for orbits
which have similar $\{r_{p},e\}$ more closely match than orbits which
have similar $\{E, L_{z}\}$ values.

\end{itemize}

This second point cannot be stressed enough, as it applies to
treatments which use the semirelativistic approximation {\it or}
Newtonian results; all approaches appear to be most
accurate when the orbits are defined by the periapse distance $r_{p}$
and eccentricity $e$ rather than by energy and angular momentum.  The
reason is that when working in flat space relating the orbit to the
curved-spacetime orbit with the same $r_{p}$ and $e$ gives much better
agreement with the curved-spacetime fluxes derived by exact methods
(Teukolsky methods) than with the fluxes from the curved-spacetime
orbit with the same $E$ and $L_z$ as the flat space orbit.  This is
one substantial improvement in accuracy (see Figure~\ref{PMComp})
which can be made for no computational cost whatsoever.

To gain further improvements, the fitting function
\erf{genfitform_arbecc} described in section~\ref{fitfunc} can be used
for only a small additional computational cost.  Using the
coefficients presented here, we have seen that it can accurately
reproduce the energy and angular momentum fluxes computed using the
semirelativistic approximation.  However, it also has more general
applicability.  Once sufficient data has been obtained by numerical
solution of the Teukolsky equation, it should be possible to derive a
good fit to that data using the same fitting ansatz.  This will provide
a more practical expression for use in astrophysical calculations.

In section~\ref{KludgeInconsis} we made use of the semirelativistic
results to estimate the inconsistency in ``kludge'' gravitational
waveforms that are being used to scope out LISA data analysis
\cite{GHK02,GG05,tev05}.  These waveforms are constructed in a similar
way to the semirelativistic fluxes described here, but the inspiral
trajectory of the particle is computed independently of the waveforms
using post-Newtonian results.  We saw that the energy content of the
gravitational waves can be as much as a factor of three greater than
the energy lost by the particle orbit.  This is an important point to
bear in mind when interpreting results computed using these
approximations.

The semirelativistic formalism presented here should find uses in
computational problems where speed is of concern (e.g., large
numerical simulations) and the role played by the central black hole
is important to the dynamics of individual particles in the problem.
Such problems of interest might include new simulations of star
cluster evolution in galactic nuclei to estimate the LISA EMRI event
rate, or supermassive black hole inspiral simulations which seek to
use interactions with stellar populations as a source of dynamical
friction to bring the large black holes into proximity.  In a
companion paper \cite{GKL1}, we use the insight gained here, in
conjunction with numerical results from solution of the Teukolsky
equation \cite{martel04}, to compute improved expressions for the
inspiral timescale of capture orbits.  The resulting expressions can
be easily included in simulations of stellar clusters
\cite{Freitag01,Freitag02,Freitag03} to improve estimates of capture
rates.

The semirelativistic approximation can also be applied to estimate energy and angular momentum fluxes from objects orbiting Kerr black holes. The procedure is more complicated due to the inclusion of spin and lack of spherical symmetry. In particular, it is not clear how to evolve the third integral of the motion, the Carter constant, for Kerr inspirals. However, by identifying Boyer-Lindquist coordinates with flat-space spherical polar coordinates and constructing the corresponding flat-space quadrupole moment tensor in the manner employed here, estimates for the energy and angular momentum fluxes from Kerr orbits may still be obtained. Preliminary results suggest that such semirelativistic estimates improve over standard post-Newtonian results \cite{ryan96} for spinning black holes as well. To construct inspirals, the angular momentum and energy fluxes can be combined with ``kludge'' approximations for the evolution of the Carter constant \cite{GHK02,GG05}. This extension to Kerr will be described in a future paper.

\acknowledgments We would like to thank E.\ Sterl Phinney for
initially suggesting this problem, and the members of the TAPIR group
at Caltech for helpful discussions during the completion of this work.
We also thank Kostas Glampedakis and Stanislav Babak for several
helpful suggestions. SLL and JRG thank the Aspen Centre for Physics for their hospitality while the manuscript was being finished. This work was supported in part by NASA grants
NAG5-12834 (JRG, DJK) and NAG5-10707 (JRG).  SLL acknowledges support
at Penn State from the Center for Gravitational Wave Physics, funded
by the NSF under cooperative agreement PHY 01-14375, as well as
support from Caltech under LISA contract number PO 1217163.

\appendix
\section{Analytic results for arbitrary eccentricity}
\subsection{Exact expressions}
\label{exactapp}
For orbits of arbitrary eccentricity it is also possible to derive
exact expressions for the loss in energy and angular momentum, which
reduce to \erf{PardeltaE} and \erf{PardeltaLz} in the parabolic case.
This is accomplished by writing the energy and angular momentum lost
as a sum of integrals of the form
\begin{equation}
   I_{n}=\int_{r_{p}}^{r_{a}} \frac{M^{n+1}\,\rmd r}{r^{n}
   \sqrt{\left(r_{a}-r\right)\left(r-r_{p}\right)\left(r-r_{-}\right)\,r}}.
   \label{intform}
\end{equation}
By considering the derivative of
$\sqrt{(r_{a}-r)(r-r_{p})(r-r_{-})r}/r^{n}$ and using results in
\cite{grad94}, we deduce
\begin{eqnarray}
   I_{n}&=&\left(
   \frac{n-1}{2n-1}\right) I_{n-1}-\frac{M((1+e)r_{p} - (3 + e^{2}) M)}
   {r_{p}^{2} (1+e)^{2}} \frac{(2n-3)}{(2n-1)} I_{n-2}   +\frac{M^{2}(1-e)((1+e)r_{p}-4 M)}{r_{p}^{3} (1+e)^{2}} 
   \left(\frac{n-2}{2n-1}\right) I_{n-3} \nonumber \\
   I_{0}&=& \frac{2 M}{r_{p}} \sqrt{\frac{(1-e)((1+e)r_{p}-4 M)}
   {(1+e)((1+e)r_{p}-2(3-e) M)}} {\bf K} \left(\sqrt{\frac{4e M}
   {((1+e)r_{p}-2(3-e) M)}}\,\right) \nonumber \\
   I_{1}&=& \frac{M((1+e)r_{p}-4 M)}{(1+e) r_{p}^{3}}
   \sqrt{\frac{(1-e)((1+e)r_{p}-4 M)}{(1+e)((1+e)r_{p}-2(3-e) M)}}
   \nonumber \\
   && \times \left[r_{p} {\bf K} \left(\sqrt{\frac{4e M}{((1+e)r_{p}-2(3-e)
   M)}}\right) -\frac{r_{p} ((1+e)r_{p}-2(3-e) M)}{(1+e)r_{p}-4 M} {\bf E}
   \left(\sqrt{\frac{4e}{((1+e)r_{p}-2(3-e) M)}}\right)\right].
   \label{genrec}
\end{eqnarray}
The functions ${\bf K}(k)$ and ${\bf E}(k)$ are the complete elliptic
integrals of the first and second kinds \erf{ellints}.  Using the
recurrence relation \erf{genrec} we can express the energy and angular
momentum lost in terms of these elliptic integrals.  We find the
expression for the energy loss to be
\begin{eqnarray}
   \frac{M}{m} \Delta E &=& -\frac{16 M^{11}}{1673196525 r_{p}^{6} 
   (1+e)^{\frac{19}{2}} \left((r_{p}-2 M)((1+e)r_{p}-2(1-e) 
   M)\right)^{\frac{5}{2}}} \nonumber \\
   && \times \,\left[ \sqrt{(1+e)\frac{r_{p}}{M}-2(3-e)} {\bf E}
   \left(\sqrt{\frac{4 e M}{((1+e)r_{p}-2(3-e) M)}}\right) 
   \,f_{1}\left(\frac{r_{p}}{M},e\right) \right. \nonumber \\
   && \left.  + \frac{(1+e)}{\sqrt{(1+e)\frac{r_{p}}{M}-2(3-e)}} \,
   {\bf K} \left( \sqrt{\frac{4 e M}{((1+e)r_{p}-2(3-e) M)}}\right)
   \,f_{2}\left(\frac{r_{p}}{M},e\right)\right]
   \label{GendeltaE}
\end{eqnarray}
where
\begin{eqnarray}
    f_{1}\left(y,e\right) = && 4608 \left( 1 - e \right) \,{\left( 1 +
    e \right) }^2\, {\left( 3 + e^2 \right) }^2\, \left( 2428691599 +
    313957879\,e^2 + 1279504693\,e^4 \right. \nonumber \\
    && \left.  + 63843717\,e^6 \right) + 192\,{\left( 1 + e \right) }^2\,
    \left( 908960573673 - 155717471796\,e^2 \right. \nonumber \\
    && \left.- 88736969547\,e^4 - 293676299040\,e^6 - 195313674237\,e^8 -
    26635698156\,e^{10} \right. \nonumber \\
    && \left.  - 346799201\,e^{12} \right) \,y - 384\,{\left( 1 + e
    \right) }^3\, \left( 336063804453 - 53956775638\,e^2 \right. \nonumber \\
    && \left.  - 33318942522\,e^4 - 92857670352\,e^6 -
    41764459155\,e^8 - 2765710514\,e^{10} \right) \,y^2  \nonumber \\
    && + 16\,{\left( 1 + e \right) }^4\, \left( 3418907055555
    - 580720618635\,e^2 - 168432860626\,e^4 \right. \nonumber \\
    && \left.  - 606890963686\,e^6 - 176495184865\,e^8 -
    3768291999\,e^{10} \right) \,y^3 \nonumber \\
    && - 32\,{\left( 1 + e \right) }^5\, \left(510454645597 -
    92175635794\,e^2 + 26432814256\,e^4 - 28250211070\,e^6 \right. \nonumber \\
    && \left.  - 5713846269\,e^8\right) \,y^4 + 4\,{\left( 1 + e \right) }^6
    \left( 1107402703901 - 174239346926\,e^2 \right. \nonumber \\
    && \left.  + 100957560852\,e^4 + 3707280110\,e^6 - 899162673\,e^8
    \right) \,y^5 \nonumber \\
    &&  - 8\,{\left( 1 + e \right) }^7\, \left( 143625217397 - 16032820010
    e^2 + 4238287541\,e^4 + 275190560\,e^6 \right) \,y^6  \nonumber \\
    && + {\left( 1 + e \right)}^8\, \left( 220627324753 - 14884378223\,e^2
    - 1210713997\,e^4 + 14138955\,e^6 \right) \,y^7  \nonumber \\
    && -8\,{\left( 1 + e \right) }^9\, \left( 2922108518 - 46504603\,e^2 -
    2407656\,e^4 \right) \,y^8 \nonumber \\
    && + 3\,{\left(1 + e \right) }^{10}\, \left( 241579935 + 6314675\,e^2 -
    149426\,e^4\right) \, y^9 \nonumber \\
    &&  + 4\,{\left( 1 + e \right) }^{11}\,
    \left( 8608805 - 48992\,e^2 \right) \,y^{10} + 2\,{\left( 1 + e
    \right) }^{12}\, \left( 1242083 - 16320\,e^2 \right) \,y^{11} \nonumber \\
    && + 184320\,{\left( 1 + e \right)}^{13}\,y^{12} +
    5120\,{\left( 1 + e \right) }^{14}\,y^{13} \nonumber
\end{eqnarray}
and
\begin{eqnarray}
    f_{2}(y,e) = &&3072\,\left( 3 - e \right) \,\left( 3 + e \right) 
    \left( 3 + e^2 \right) \left( 7286074797 - 3299041125\,e^2 +
    792940362\,e^4 \right.  \nonumber \\
    &&  \left.  - 1366777698\,e^6 - 369698151\,e^8
    - 5932745\,e^{10} \right) - 384\,\left( 1 + e \right) \left(
    2989180413711 \right. \nonumber \\
    && \left.- 583867932642\,e^2 - 131661872359\,e^4 - 419423580924\,e^6 -
    194293515951\,e^8 \right. \nonumber \\
    && \left.- 3390301442\,e^{10} + 1353430119\,e^{12} \right) \,{y} +
    64\,{\left( 1 + e \right) }^2\, \left( 14825178681327 \right. \nonumber \\
    && \left.  - 2675442646782\,e^2 - 728511901515\,e^4- 1837874368340\,e^6 
    - 591999524567\,e^8 \right. \nonumber \\
    && \left.  - 1856757710\,e^{10} + 841581651\,e^{12}\right) \,{y}^2 -
    32\,{\left( 1 + e \right) }^3\, \left(14292163934541 \right. \nonumber \\
    && \left.  - 2666166422089\,e^2 - 522582885086\,e^4 - 1347373382962\,e^6 -
    307066297439\,e^8 \right. \nonumber \\
    && \left.- 1675056789\,e^{10} \right) \,{y}^3 + 16\,{\left( 1 + e
    \right) }^4\,\left( 9557748374919 - 1917809903861\,e^2 \right. \nonumber \\
    && \left.  - 24258045506\,e^4 - 511875047746\,e^6 - 86779453317\,e^8 -
    462078345\,e^{10} \right) \,{y}^4 \nonumber \\
    && - 8\,{\left( 1 + e \right) }^5\, \left(5390797838491 -
    990602472036\,e^2 + 161182699002\,e^4 \right. \nonumber \\
    && \left.  - 89978894004\,e^6 - 11363685245\,e^8 \right) \,{y}^5 +
    4\,{\left( 1 + e \right) }^6\, \left( 2857676457065 \right. \nonumber \\
    && \left.  - 351292910556\,e^2 + 79840371470\,e^4 - 2670080940\,e^6 -
    463345647\,e^8 \right) \,{y}^6 \nonumber \\
    && -2\,{\left( 1 + e \right) }^7\, \left( 1249768416047 - 79903103833\,e^2
    + 12179840133\,e^4 \right. \nonumber \\
    && \left.  + 482157413\,e^6 \right) \,{y}^7 + {\left( 1 + e \right) }^8
    \left(363565648057 - 10040939153\,e^2 - 318841465\,e^4 \right. \nonumber \\
    &&  \left.  + 14611473\,e^6 \right) \,{y}^8 - 2\,{\left(
    1 + e \right) }^9\, \left( 13862653487 - 100645509\,e^2 -
    11015842\,e^4 \right) \,{y}^9 \nonumber \\
    && + {\left(1 + e \right) }^{10}\, \left( 518128485 + 16345427\,e^2 -
    421398\,e^4\right) \,{y}^{10} \nonumber \\
    && + 16\,{\left( 1 + e\right) }^{11}\, \left( 1220639 - 13448\,e^2
    \right) \,{y}^{11} +2\,{\left( 1 + e \right) }^{12}\, \left( 689123 -
    18880\,e^2 \right)\,{y}^{12}  \nonumber \\
    && +153600\,{\left( 1 + e \right)}^{13}\,{y}^{13} + 5120
    {\left( 1 + e \right) }^{14}\,{y}^{14}
    \nonumber
\end{eqnarray}
The angular momentum lost is similarly given by
\begin{eqnarray}
   \frac{\rmd L_{z}}{m} &=& -
   \frac{16\,M^{\frac{15}{2}}}{24249225 \left(1+e\right)^{\frac{13}{2}}
   r_{p}^{\frac{7}{2}}\,(r_{p}-2\,M)^{2}\,\left((1+e)r_{p}-2(1-e) 
   M\right)^{2}} \nonumber \\
   && \times \left[\sqrt{(1+e)\frac{r_{p}}{M}-2(3-e)}\,{\bf E}
   \left(\sqrt{\frac{4\,e\,M}{((1+e)r_{p}-2(3-e)\,M)}}\right) \,
   g_{1}\left(\frac{r_{p}}{M},e\right) \right. \nonumber \\
    && \left.  + \frac{(1+e)}{\sqrt{(1+e)\frac{r_{p}}{M}-2(3-e)}}
    {\bf K} \left(\sqrt{\frac{4\,e\,}{((1+e)r_{p}-2(3-e)\,M)}}\right)
   \,g_{2}\left(\frac{r_{p}}{M},e\right)\right]
   \label{GendeltaL}
\end{eqnarray}
where
\begin{eqnarray}
    g_{1}(y,e)= &&169728\,\left( 1 - e \right) \,{\left( 1 + e
    \right) }^2\, \left( 279297 + 219897\,e^2 + 106299\,e^4 + 9611\,e^6
    \right) \nonumber \\
    && - 384\,{\left( 1 + e \right)}^2\, \left( 192524061 - 13847615\,e^2
    - 36165965\,e^4 - 20710173\,e^6 - 588532\,e^8 \right) \,y  \nonumber \\
    && + 192\,{\left( 1 + e \right) }^3\, \left( 235976417 + 13109547\,e^2 -
    3369705\,e^4 - 3292707\,e^6 \right) \,y^2  \nonumber \\
    && - 16\,{\left( 1 + e \right) }^4\, \left( 813592799 +
    112906199\,e^2 + 53843933\,e^4 + 602061\,e^6 \right) \,y^3 \nonumber \\
    && + 16\,{\left( 1 + e \right) }^5\, \left( 87491089 + 7247482\,e^2 +
    4608349\,e^4 \right) \, y^4 + 8\,{\left( 1 + e \right) }^6\, \left(
    9580616 \right. \nonumber \\
    && \left.+ 6179243\,e^2 - 92047\,e^4 \right) \,y^5 - 4\,{\left( 1 + e
    \right) }^7\, \left( 3760123 + 272087\,e^2 \right) \,y^6 \nonumber \\
    && - {\left( 1 + e \right) }^8\, \left( 1168355 -
    35347\,e^2 \right) \,y^7 - 71792\,{\left( 1 + e \right) }^9\,y^8 -
    4120\,{\left( 1 + e \right) }^{10}\,y^9 
    \nonumber
\end{eqnarray}
and
\begin{eqnarray}
   g_{2}(y,e) = && 339456\,\left( 3 - e \right) \,\left( 3 + e \right) \,
   \left( 93099 - 10213\,e^2 - 18155\,e^4 - 10551\,e^6 - 420\,e^8 \right)
   \nonumber \\
    && - 1536\,\left( 1 + e \right) \, \left( 319648410 - 35712133\,e^2
    - 33099777\,e^4 - 11272311\,e^6 + 457187\,e^8 \right) \,{y} \nonumber \\
    && + 128\,{\left(1 + e \right) }^2\, \left( 2706209781 - 45415294\,e^2 -
    103634296\,e^4 - 34056010\,e^6 - 130293\,e^8 \right) \,{y}^2  \nonumber \\
    && - 32\,{\left( 1 + e \right) }^3\, \left( 3895435659 +
    212168215\,e^2 + 4641265\,e^4 - 15197651\,e^6 \right) \,{y}^3 \nonumber \\
    && + 16\,{\left( 1 + e \right) }^4\, \left(
    1396737473 + 123722895\,e^2 + 27602127\,e^4 - 465119\,e^6 \right)
    \,{y}^4 \nonumber \\
    && - 16\,{\left( 1 + e \right)
    }^5\, \left( 78148621 + 3035912\,e^2 + 3130827\,e^4 \right) \, {y}^5
     \nonumber \\
    && - 16\,{\left( 1 + e \right) }^6\, \left( 8005570 + 1485159\,e^2 -
    47943\,e^4 \right) \,{y}^6 + 2\,{\left( 1 + e \right) }^7\, \left(
    4015181 + 601959\,e^2 \right) \,{y}^7 \nonumber \\
    && + {\left( 1 + e \right) }^8\, \left( 737603 - 39467\,e^2
    \right) \,{y}^8 + 47072\,{\left( 1 + e \right) }^9\,{y}^9 +
    4120\,{\left( 1 + e \right) }^{10}\,{y}^{10}
    \nonumber
\end{eqnarray}
The limit $r_{p} \rightarrow \infty$ corresponds to the argument of the elliptic integrals approaching zero.  Using series expansions of the elliptic integrals about $k=0$ \cite{abram64}, we find
\begin{eqnarray}
   \frac{M}{m}\,\Delta E &\approx& -\frac{64\,\pi}{5} \,
   \frac{1}{\left(1+e\right)^{\frac{7}{2}}}\,\left(1+\frac{73}{24}
   e^{2}+\frac{37}{96}\,e^{4}\right)\,\left(
   \frac{r_{p}}{M}\right)^{-\frac{7}{2}} \nonumber \\
    && \hspace{0.5in} - \frac{64\,\pi}{5} \frac{1}{\left(1+e\right)^{\frac{9}{2}}}
    \left(1+\frac{31}{8}\,e^{2}+\frac{65}{32}\,e^{4}+\frac{1}{6}
    e^{6}\right) \left(\frac{r_{p}}{M}\right)^{-\frac{9}{2}}  +\,O\left(\left(\frac{r_{p}}{M}\right)^{-\frac{11}{2}}\right)
    \nonumber \\
   \frac{\Delta L_{z}}{m} &\approx& -\frac{64\,\pi}{5}\,
   \frac{1}{\left(1+e\right)^{2}}\, \left(1+\frac{7}{8}\,e^{2}\right)\,\,
   \left(\frac{r_{p}}{M}\right)^{-2} \nonumber \\
    && \hspace{0.5in} - \frac{192\,\pi}{5}\,
   \frac{1}{\left(1+e\right)^{3}}\,\left(1+\frac{35}{24}
   \,e^{2}+\frac{1}{4}\, e^{4}\right)\,\,
   \left(\frac{r_{p}}{M}\right)^{-3}  + O\left(\left(\frac{r_{p}}{M}\right)^{-4}\right)\ .
   \label{genasymexp}
\end{eqnarray}
The leading order terms agree with the Keplerian results
\erf{PMdE}--\erf{PMdL} \cite{PM63}, as expected.  In the limit $r_{p}
\rightarrow r_{UCO} = 2(3+e)/(1+e)$, the argument of the elliptic
integrals approaches $1$.  The elliptic integrals diverge
logarithmically in this limit, and we may expand them as in equations
\erf{asymK} and \erf{asymE}.

On substitution of these expansions into
\erf{GendeltaE}--\erf{GendeltaL}, we find the asymptotic form of
$\Delta E$ and $\Delta L_{z}$ to be
\begin{equation}
   \frac{M}{m}\,\Delta X \approx p_{X}\,\ln{\left(\frac{r_{p}}{M}
   -\frac{2(3+e)}{1+e}\right)}\,+\,q_{X} + O\left(\left(\frac{r_{p}}{M}-\frac{2(3+e)}{1+e}\right)
   \ln{\left(\frac{r_{p}}{M}-\frac{2(3+e)}{1+e}\right)}\right)
   \label{Genasymth}
\end{equation}
As before, $X$ refers to either $E$ or $L_{z}/M$.  The coefficients
$p_{X}$ and $q_{X}$ are functions of eccentricity
\begin{eqnarray}
	p_{E} &=& \frac{4\,{\left( 1 + e \right) }^
   {\frac{7}{2}}}{5\,{\sqrt{e}}\, {\left( 3 + e \right) }^3}
   \label{pE} \\
    q_{E} &=& 4\,{\sqrt{e}}\, \left( 126493657290 +
    548139181590\,e + 1030019780790\,{e}^2 \right.\nonumber \\
   && \left.  + 1139255611065\,{e}^3 + 838466930873\,{e}^4 +
   401719467929\,{e}^5 + 98700067049\,{e}^6\right.\nonumber \\
   && \left.  + 6236043751\,{e}^7 + 2856045401\,{e}^8 - 177251547\,{e}^9
   - 1203124043\,{e}^{10} \right. \nonumber \\
   && \left. + 316812556\,{e}^{11} + 109455696\,{e}^{12} -
   88995328\,{e}^{13} \right)/\left( 1673196525\,{\left( 1 + e
   \right) }^ {\frac{5}{2}}\,{\left( 3 + e \right) }^6\right)\nonumber \\
   && - \frac{4\,{\left( 1 + e\right) }^ {\frac{7}{2}}\,\left( \ln (64)
   - \ln (1 + e)+ln(e) \right)}{5\, {\sqrt{e}}\, {\left( 3 + e
   \right) }^3}
   \label{qE} \\
 p_{L_{z}} &=& \frac{8\,\sqrt{2}\,(1+e)^{2}}{5\,{\left(3 +
   e\right)}^{\frac{3}{2}}\, \sqrt{e}}
   \label{pLz} \\
    q_{L_{z}} &=& \frac{16\,\sqrt{2}\,(1+e)^{2}}{24249225\,
    {\left( 3 + e\right)}^{\frac{3}{2}}\,\sqrt{e}} 
    \left( \frac{e}{{\left( 1 + e \right) }^4\, {\left( 3 + e \right) }^2}
    \left( 174594420 + 523783260\,e \right. \right. \nonumber \\
    && \left. \left. + 557732175\,{e}^2 + 241337525\,{e}^3 + 44249062\,{e}^4
    + 11244922\,{e}^5 - 2993241\,{e}^6  - 1809123\,{e}^7 \right.
    \right. \nonumber \\
    && \left. \left. + 1328784\,{e}^8 -
    172744\,{e}^9 \right) - \frac{4849845\,\left(
    6\ln (2) - \ln (1 + e)+\ln(e) \right) }{2} \right)
    \label{qLz}
\end{eqnarray}

\subsection{Fitting functions}
\label{fitapp}
We can use the exact expressions \erf{GendeltaE}--\erf{GendeltaL} to
derive fitting functions to approximate our results.  Following the
same argument used in the parabolic case, a functional form like
\erf{dEfitform} should capture the main features of the problem, but
the coefficients are now functions of eccentricity, and we replace the
parabolic value of the UCO -- $4\,M$ -- with the value appropriate to
other eccentricities.  The general ansatz is
\begin{eqnarray}
   \frac{M}{m}\Delta E &=& \left(\sum_{n=0}^{N}
   A^{E}_{n}(e)\left(\frac{M((1+e)r_{p}-2(3+e)M)}
   {(1+e) r_{p}^{2}}\right)^{n} \right) \nonumber \\
   && \times \cosh^{-1}\left[1+B_{0}^{E} 
   \left(\frac{2(3+e) M}{(1+e) r_{p}}\right)^{N_{E}-1} 
   \frac{(1+e) M}{(1+e) r_{p}-2(3+e) M}\right] \nonumber \\
   && + \frac{M^{\frac{N_{E}}{2}} ((1+e) r_{p}-2(3+e) M)}{(1+e)
   r_{p}^{1+\frac{N_{E}}{2}}} \sum_{n=0}^{N} C^{E}_{n}
   \left(\frac{M ((1+e) r_{p}-2(3+e) M)}{(1+e) r_{p}^{2}}\right)^{n}
    \nonumber \\
   && + \frac{M^{1+\frac{N_{E}}{2}} ((1+e) r_{p}-2(3+e) M)}{(1+e)
   r_{p}^{2+\frac{N_{E}}{2}}} \sum_{n=0}^{N-1} B^{E}_{n+1}
   \left(\frac{M ((1+e) r_{p}-2(3+e) M)}{(1+e) r_{p}^{2}}\right)^{n}
   \label{genfitform_arbecc}
\end{eqnarray}
Successive terms in the fit are given by matching consecutive orders
in an expansion about $r_{p}=2(3+e)/(1+e)$ and as
$r_{p}\rightarrow\infty$ in the way described in section~\ref{fitfunc}
for the parabolic case.  To illustrate, the lowest order ($N=0$)
expansion coefficients may be determined from the $\Delta E$ and
$\Delta L_{z}$ expansions \erf{genasymexp}--\erf{qLz} as follows
\begin{eqnarray}
   A^{X}_{0}(e)&=& -p_{X}(e), \qquad B_{0}^{X}(e) =
   \frac{1}{2} \exp\left(\frac{q_{X}(e)}{A_{X}(e)}\right),\nonumber \\
   C^{E}_{0}(e) &=& -\frac{64 \pi}{5}  
   \frac{1}{\left(1+e\right)^{\frac{7}{2}}} \left(1+\frac{73}{24}
   e^{2}+\frac{37}{96} e^{4}\right)-A^{E}_{0}(e) \sqrt{2 B_{0}^{E}(e)}
   \left(\frac{2(3+e)}{(1+e)}\right)^{3},\nonumber \\
   C^{L_{z}}_{0}(e) &=& -\frac{64 \pi}{5} \frac{1}{\left(1+e\right)^{2}} 
   \left(1+\frac{7}{8} e^{2}\right)-A^{L_{z}}_{0}(e) \sqrt{2 B_{0}^{L_{z}}(e)}
   \left(\frac{2(3+e)}{(1+e)}\right)^{\frac{3}{2}}\label{asymfit3}
\end{eqnarray}
Our main focus is on orbits that are nearly parabolic, with $e \approx
1$.  We therefore expand these expressions about $e=1$ to obtain
\begin{eqnarray}
   A^{E}_{0}(e) &=& -\frac{1}{5 \sqrt{2}} +
   \frac{1}{10 {\sqrt{2}}} (1-e) - \frac{1}{160 {\sqrt{2}}}  
   (1-e)^{2} + O\left((1-e)^{3}\right) \nonumber \\
   B_{0}^{E}(e) &=& 0.752091 - 0.0949439  (1-e) + 0.0918458
   (1-e)^{2} + O\left((1-e)^{3}\right) \nonumber \\
   C_{0}^{E}(e)&=& -4.63464 +1.63944 (1-e) + 0.327505  (1-e)^{2} +
   O\left((1-e)^{3}\right) \nonumber \\
   A_{0}^{L_{z}}(e) &=& -\frac{4\sqrt{2}}{5} + \frac{1}{5 {\sqrt{2}}}
   (1-e) - \frac{7}{80 {\sqrt{2}}} (1-e)^{2} +
   O\left((1-e)^{3}\right) \nonumber \\
   B_{0}^{L_{z}}(e) &=& 1.31899 - 0.126207   (1-e) + 0.392812   
   (1-e)^{2} +
   O\left((1-e)^{3}\right) \nonumber \\
   C_{0}^{L_{z}}(e)&=& -4.1491 + 1.71517  (1-e) -0.128645   (1-e)^{2}
   + O\left((1-e)^{3}\right)
   \label{fitcoeffs}
\end{eqnarray}

The expansion of the $B_{0}^{X}$'s and $C_{0}^{X}$'s may also be
written down precisely.  However, the expressions are extremely
complicated, which is why the numerical values of these coefficients
have been quoted instead.  Higher order
fitting functions may be obtained by matching more terms in the
expansions of $\Delta E$ and $\Delta L_{z}$, as described earlier. For 
completeness, we quote here the remaining coefficients of the $N=2$ 
fit, once again expanded to quadratic order about the parabolic case
\begin{eqnarray}
   A_{1}^{E}(e) &=& -0.282843   (1-e) + 0.0353553   (1-e)^{2} +
     O\left((1-e)^{3}\right) \nonumber \\
   B_{1}^{E}(e) &=& -103.215 +39.6287   (1-e) + 38.3325   (1-e)^{2} +
     O\left((1-e)^{3}\right) \nonumber \\
   C_{1}^{E}(e) &=& 69.1683 -0.682028    (1-e) -28.7945   (1-e)^{2} +
     O\left((1-e)^{3}\right) \nonumber \\
   A_{2}^{E}(e) &=& -1.20797 - 2.31872   (1-e) - 2.15134   (1-e)^{2} +
     O\left((1-e)^{3}\right) \nonumber \\
   B_{2}^{E}(e) &=& 727.515 + 1570.89   (1-e) + 1139.13   (1-e)^{2} +
     O\left((1-e)^{3}\right) \nonumber \\
   C_{2}^{E}(e) &=& -439.378 - 1223.38   (1-e) -862.812   (1-e)^{2} +
     O\left((1-e)^{3}\right) \nonumber \\
   A_{1}^{L_{z}}(e) &=& -0.565685   (1-e) + 0.494975   (1-e)^{2} +
     O\left((1-e)^{3}\right) \nonumber \\
   B_{1}^{L_{z}}(e) &=& -53.4491 +4.38709  (1-e) + 0.469838   
     (1-e)^{2} + O\left((1-e)^{3}\right) \nonumber \\
   C_{1}^{L_{z}}(e) &=& 25.4129 +16.7694   (1-e) - 7.06419   (1-e)^{2} 
     + O\left((1-e)^{3}\right) \nonumber \\
   A_{2}^{L_{z}}(e) &=& 3.9598   (1-e) - 4.80833   (1-e)^{2} +
     O\left((1-e)^{3}\right) \nonumber \\
   B_{2}^{L_{z}}(e) &=& 29.7857 + 167.281  (1-e) + 66.0607   (1-e)^{2} 
     + O\left((1-e)^{3}\right) \nonumber \\
   C_{2}^{L_{z}}(e) &=&  15.1726 - 131.512   (1-e) - 26.8611   
     (1-e)^{2} + O\left((1-e)^{3}\right)
\end{eqnarray}

\section{Erratum --- Phys. Rev. D {\bf 74} 109901 (2006)}
\label{erratum}
\begin{center}\large{{\bf Erratum: Semi-relativistic approximation to gravitational radiation from
encounters with non-spinning black holes\\ $[$Phys. Rev. D 72, 084009 (2005)$]$}}
\end{center}

\begin{center}
Jonathan R Gair, Daniel J Kennefick, Shane L Larson
\end{center}

In Section~III.E on hyperbolic orbits we stated ``The energy and angular momentum lost
to gravitational waves by a particle on a hyperbolic orbit are given by the same
Eqs. (A3)--(A4) that apply to bound orbits, just by
inserting $e>1$ consistent with the definitions (12) and (13)''. In fact, the generalisation to hyperbolic orbits is more complicated, since the integrals are now incomplete, and $I_2$ defined by equation (A1) changes since a boundary term
no longer cancels. The correct generalisation to $e>1$ requires modifying expressions (A3) and (A4) as follows
\begin{itemize}
\item Replace {\bf K}($k$) by {\bf F}($\phi$,$k$), an incomplete elliptic integral of the first kind, where
\begin{eqnarray}
k^2=\frac{4eM}{(1+e)r_p-2(3-e)M}, \qquad \phi = \sin^{-1}\left(\sqrt{\frac{(1+e)((1+e)r_p-2(3-e)M)}
{2e((1+e)r_p-4M)}}\right)\nonumber
\end{eqnarray}
\item Replace {\bf E}($k$) by {\bf E}($\phi$,$k$) - ${\bf E_H}$($r_p$,$e$), where {\bf E}($\phi$,$k$) is the incomplete elliptic integral of the second kind and\begin{eqnarray}
{\bf E_H} (r_p,e) = 2\,M\,\sqrt{\frac{e^2-1}{((1+e)r_p-2(3-e)M)\,((1+e)r_p-4M)}}\nonumber
\end{eqnarray}
\item Include an extra term, $(1+e)\,\sqrt{e^2-1}\,f_3(r_p/M,e)/\sqrt{(1+e)r_p/M-4}$, inside the square bracket of equation (A3), and a term,  $(1+e)\,\sqrt{e^2-1}\,g_3(r_p/M,e)/\sqrt{(1+e)r_p/M-4}$, inside the square bracket of equation (A4). The new functions are given by
\begin{eqnarray}
f_3(y,e) &=& 5120 (e+1)^{13} y^{13}+186880 (e-1) (e+1)^{12}
   y^{12}-2 (e+1)^{11} \left(15040
   e^2-1294563\right) y^{11} \nonumber \\ && -(e+1)^{10} \left(89728
   e^2-36202743\right) y^{10}-2 (e+1)^9
   \left(229899 e^4-10685662 e^2 -375412235\right)
   y^9  \nonumber \\ &&  +(e+1)^8 \left(19031885 e^4-2703503366
   e^2-25681442087\right) y^8 +4 (e+1)^7
   \left(3474054 e^6-1305675539 e^4 \right.  \nonumber \\ &&  \left. +3107359416   e^2+59447608277\right) y^7  -4 (e+1)^6
   \left(865886773 e^6-22415704847 e^4-30407605409
   e^2 \right.  \nonumber \\ &&  \left. +296614649595\right) y^6  +16 (e+1)^5
   \left(534461702 e^8+11533033897 e^6-22057127975
   e^4-76603770053 e^2 \right.  \nonumber \\ && \left. +272389989629\right) y^5-16
   (e+1)^4 \left(17178999909 e^8+162890646772
   e^6+59295125158 e^4 \right.  \nonumber \\ && \left. -287223884988
   e^2+971850401469\right) y^4 +64 (e+1)^3
   \left(603682818 e^{10}+45406272911 e^8 \right.  \nonumber \\ && \left. +243208507040
   e^6+155132151938 e^4-157983734058
   e^2+799957785207\right) y^3  \nonumber \\ && -64(e+1)^2
   \left(8058145731 e^{10}+210515698689 e^8+706986172382
   e^6+411976993282 e^4 \right.  \nonumber \\ && \left. -263668934513
   e^2+1861696008653\right) y^2+256 (e+1)
   \left(35596470 e^{12}+8535177147 e^{10} \right.  \nonumber \\ && \left. +111317681695
   e^8+249751373150 e^6+118284328500 e^4-88270736617
   e^2+620968567495\right) y  \nonumber \\ && -2048 \left(17798235
   e^{12}+1445881344 e^{10}+11131231123 e^8+17466667640
   e^6+6464885073 e^4 \right.  \nonumber \\ && \left. -7838537848 e^2+44213644993\right)\nonumber \\
g_3(y,e) &=& -4120 (e+1)^9 y^9-73852 (e+1)^8 y^8+
   (e+1)^7 \left(33287 e^2-1214551\right) y^7 +4
   (e+1)^6 \left(929703 e^2 \right.  \nonumber \\ &&  \left. +3916957\right) y^6+16
   (e+1)^5 \left(257955 e^4+2125555 e^2 -10535542\right)
   y^5-16  (e+1)^4 \left(6209867 e^4 \right.  \nonumber \\ &&  \left. +36422010
   e^2-127264229\right) y^4 -64 (e+1)^3 \left(532495
   e^6+749107 e^4-34605291 e^2+203477737\right) y^3  \nonumber \\ &&  +128
   (e+1)^2 \left(7591871 e^6+40714475 e^4-24139707
   e^2+328231985\right) y^2   -256(e+1) \left(278460
e^8 \right.  \nonumber \\ &&  \left. +22737761 e^6+79099653 e^4-7768285
   e^2+262081211\right) y  +226304 \left(e^2+3\right)
   \left(1260 e^6+40039 e^4 \right.  \nonumber \\ &&  \left. -23378 e^2+62719\right) \nonumber
\end{eqnarray}
\end{itemize}
The results shown in Figure 8 were computed by numerical integration of the
fluxes, and are therefore correct inspite of this error. This erratum has been published as {\it Phys. Rev.} D {\bf 74} 109901.

\end{document}